\documentclass[draft]{agujournal2019}
\usepackage{url} 
\usepackage{lineno}
\usepackage[finalnew, ignoremode]{trackchanges} 
\usepackage{soul}
\usepackage[T1]{fontenc}
\soulregister\ref7
\soulregister\cite7
\addeditor{DW}

\newcommand{\av}[1]{\langle #1\rangle}
\usepackage{bm}
\newcommand* {\vek}[1]{{\ensuremath{\bm{\mathrm{#1}}}}}

\newcommand{\sfn}{$S^{fn}$}
\newcommand{\sfns}{$S^{fn}$s}
\draftfalse

\journalname{Space Weather}

\begin{document}

%
%

\title{Exploring the potential of neural networks to predict statistics of solar wind turbulence}

%
%




\authors{Daniel Wrench\affil{1}, Tulasi N. Parashar\affil{1}, Ritesh K. Singh\affil{2}, Marcus Frean\affil{1}, Ramesh Rayudu\affil{1}}

\affiliation{1}{Victoria University of Wellington, Kelburn, Wellington, NZ 6012}
\affiliation{2}{Department of Physical Sciences, Indian Institute of Science Education and Research Kolkata, Mohanpur, 741246, India}





\correspondingauthor{Daniel Wrench}{daniel.wrench@vuw.ac.nz}




\begin{keypoints}

\item Small neural networks are able to predict structure functions for sparse solar wind time series in a limited sense.

\item A network with only 20 hidden neurons statistically outperforms (in terms of MSE) simple imputation techniques for high ($>$50\%) data loss.

\item More work is needed to generalize the model's architecture to improve performance and increase applicability to other systems.

\end{keypoints}

%
%

%
%


\begin{abstract}
Time series datasets often have missing or corrupted entries, which need to be \change[DW]{ignored}{handled} in subsequent data analysis. For example, in the context of space physics, calibration issues, satellite telemetry issues, and unexpected events can make parts of a time series unusable. This causes problems for understanding the dynamics of the heliosphere and space weather environment. Various approaches exist to tackle this problem, including mean/median imputation, linear interpolation, and autoregressive modeling. Here we study the utility of artificial neural networks (ANNs) to predict statistics of sparse time series. Our focus is not on time series {\em prediction} but on gleaning the best possible information about the statistical behaviour of the system. As an example application, we focus on the structure functions of turbulent time series measured in the solar wind. Using a dataset with artificial gaps, a neural network is trained to predict second-order structure functions and then tested on an unseen dataset to quantify its performance. A small feedforward ANN, with only 20 hidden neurons, can predict the large-scale fluctuation amplitudes better than mean imputation or linear interpolation when the percentage of missing data is high. Although they perform worse than the other methods when it comes to capturing both the shape and fluctuation amplitude together, their performance is better in a statistical sense for large fractions of missing data. Caveats regarding their utility, the optimisation procedure, and potential future improvements are discussed.
\end{abstract}

\section*{Plain Language Summary}
We explore the utility of machine learning to predict statistics of a turbulent system such as the solar wind, in cases involving large data gaps. It is shown that simple artificial neural networks (ANNs) are good at estimating large-scale features of second-order structure functions even for very large amounts of missing data. However, these simple ANNs are limited in estimating other features of the structure functions, such as inner and outer scales, and the inertial range slope. More sophisticated methods are required to describe such features. Developing such a method is key to improving space weather models (e.g. the functions that couple solar wind parameters to space weather) in the face of incomplete data.

%
%

\section{Introduction}

Analyses of real-world time series are often hindered by incomplete datasets. This is very common for physiological, environmental, astronomical, and heliospheric time series \add[DW]{\cite{Rehfeld2011}}. The instrumentation used to take measurements may be prone to failure, or variations in the environment itself may preclude data collection for certain periods. For example, time series of sea level and wave height based on radio signals are commonly incomplete due to radio interference, airborne seawater spray, and the loss of line-of-sight caused by large waves \cite{Makarynskyy2005}. In physiology, recordings of blood flow and other processes are often contaminated with artifacts due to movement of the subject and improper interfacing with sensors \cite{Pavlova2019}, and removal of these leaves gaps in the series. Ground-based astronomical observations are affected by cloud cover and the maintenance and malfunction of instruments. In the case of \textit{in situ} measurements of the solar wind ---vital for models of space weather--- incomplete time series result from calibration, instrumentation, and telemetry issues \remove[DW]{\cite{Rehfeld2011}}. Telemetry is a particular issue for the two Voyager spacecraft, which must align their data transmissions with NASA's ground-based communication facilities, the Deep Space Network \cite{Ludwig2016, Gallana2016}.

Discontinuity in time series data represents a loss of information, affecting the statistics and in turn polluting predictions. This includes significant effects on frequency-domain (spectral) and scale-domain analysis. An example is `spectral inheritance', in which the gaps contaminate the rest of the data in the form of ``spurious periodicities arising from the spectral properties of the sets of gaps'' \cite{Frick1998, Gallana2016}. More generally, data gaps result in dirty spectra, which lead to poor estimation of power, particularly at high frequencies \cite{Munteanu2016}. In radio and gamma-ray astronomy, this causes issues for calculating the periodicity of stellar objects 
\cite{VanderPlas2018}. In heliophysics it hinders our understanding of the spectral properties of turbulence \cite{Fraternale2019}.
For example, \citeA{Gallana2016} highlighted the importance of identifying new techniques to extract useful spectral information from high sparsity solar wind datasets.

This is the motivation for the present study. Notably, missing data not only affect measurements taken at great distance from the Sun, but also at the near-Earth plasma environment, a region from which data is essential to developing magnetic field models for space weather research. OMNI (near-Earth magnetic field) data has 20\% of plasma parameter data missing between 2000 and 2019, and 8\% of interplanetary magnetic field (IMF) measurements over the same period. These gaps are particularly significant during storm events \cite{Kobayashi2021}. Intervals prior to 1995 are of even worse quality due to fewer spacecraft being in operation \cite{Qin2007}, leading to some gaps of more than 150 hours \cite{Finch2007}. As well as affecting spectral analysis, such gaps have been shown to introduce considerable errors into combinations of near-Earth solar wind parameters known as `coupling functions' \cite{Finch2007, Lockwood2019}. This data quality issue also presents a problem for machine learning models, which are increasingly popular algorithms for space weather analysis but generally require continuous inputs \cite{Sharpe1995, Kobayashi2021}. Therefore, the development of a technique, or series of techniques, to be able to utilise such poor-quality datasets is essential to develop a more comprehensive space climatology, and in turn, make more accurate predictions of space weather events.

Many different methods have been explored to deal with this issue of spectral estimation from a time series that has gaps. A significant amount of literature has been dedicated to estimating the power spectra and periodicities of a gapped signal. We find that these methods can be grouped into two broad categories:

\begin{enumerate}
    \item Interpolation of missing values, followed by spectral estimation from the reconstructed signal
    \item Spectral estimation directly from the dataset with gaps
\end{enumerate}

The first category of techniques is regularly used in the space plasma literature \cite[and references therein]{Lockwood2019}. Often, intervals of data that contain large gaps are simply excluded from analysis. For example, \citeA{wu_2013} and \citeA{ChenApJS20} removed \add[DW]{intervals with} gaps larger than 5\% and 1\% respectively. The remaining small gaps are typically filled using linear interpolation \cite{burlaga_91, podesta2007}. Linear interpolation has been shown to perform better than a range of other interpolation methods for estimating the values in small gaps in OMNI time series, as measured by root mean-squared error (RMSE) and $R^2$ \cite{Kobayashi2021}, where $R^2$ measures the association between the interpolated values and the true values. However, linear interpolation amounts to strong smoothing of part of the signal, which results in a loss of information at high frequencies \cite{Frick1998}. Because this effect becomes worse with increasing data loss, linear interpolation is only feasible for relatively small gaps \cite{bavassano, chen_2002}. Furthermore, by excluding large segments of the data to avoid the gaps, a considerable amount of information about the system is lost. For this reason, interpolation methods that are more consistent with the spectral content of the observed data segments have also been used. 

For example, interpolation of sparse signals has also been achieved by modelling the signal as a stochastic process (specifically, that of fractional Brownian motion), and then further defining the process as a multi-point ``bridge'' between the prescribed (observed) measurements \cite{Friedrich2020}. A strategy for identifying the optimal Hurst exponent required by the fractional Brownian motion algorithm was proposed and tested by reconstructing velocity field measurements from a superfluid helium experiment.

Singular spectrum analysis (SSA) is a non-parametric algorithm used for forecasting from gapped time series in a number of fields, including heliophysics \cite{Schoellhamer2001, Kondrashov2010}. SSA involves reconstructing a signal from its principal components, and its benefits are that it requires no prior knowledge of the periodicities in the data, and it accounts for noise in the signal. However, the technique is especially sensitive to increasing gap sizes: the RMSE was shown to increase significantly in a study investigating data gaps in soil respiration data \cite{Zhao2020}. SSA has been used to fill the significant gaps in solar wind data pre-1995 by \citeA{Kondrashov2010}. This study was able to reconstruct the measurements with realistic variability across both storm and quiet conditions. However, the study also made use of simultaneous continuous measurements of geomagnetic indices, and such proxies are not always available.

ARIMA models are the standard models for forecasting time series, and these can be fitted to non-uniformly sampled data using a maximum likelihood technique \cite{Harvey1984, Broersen2006}. This has been shown to result in much better estimation of time series parameters such as level, error variance, and slope of the time series, compared to simple mean imputation and linear interpolation \cite{Velicer2005}. A similar method of finding the best ARIMA model order based on maximising entropy has been applied to solar oscillation data \cite{Brown1990}. Starting with some assumptions about the typical gap-lengths and the noise in the signal, the authors were able to reproduce unique spectral features.

Neural networks, a prominent algorithm from machine learning, have also been used to fill gaps in time series. Specifically, a simple feed-forward neural network was found to accurately reproduce simulated stochastic processes and fill gaps that matched the original power spectrum with up to 50\% missing data \cite{Comerford2015}. Generative adversarial networks \cite{Luo2018} and convolutional neural networks \cite{jang_2020} have also been used to impute missing intervals.

A comprehensive study of dealing with large data gaps in solar wind data used a combination of techniques to recover the spectrum from Voyager datasets \cite{Gallana2016}. This compared  Fourier transforms of gap-free subsets; Fourier transforms of the correlation function of the data, with and without linear interpolation; maximum likelihood recovery; and compressed sensing spectral estimation. All of these methods, apart from compressed sensing, fall into the first category of gapped estimation techniques. Ultimately, this work was able to determine spectra over a very large range of frequencies for data with up to 50\% missing and thereby extract information on various turbulent features.

Moving now to the second category of spectral estimation methods, a continuous wavelet transform method has been used to perform spectral estimation directly from a gapped signal \cite{Frick1998}. Of direct relevance to our work, this technique has been applied to magnetic field time series in the solar system by \citeA{magrini_2017} and \citeA{deSouzaEcher2021}. In the first study, the wavelet method was compared with two polynomial interpolation methods for spectral analysis of artificially-gapped OMNI data. It was found that all techniques perform satisfactorily for small gaps, but the wavelet method better estimates the energy of the signal at certain scales for large gaps. In the second study, the wavelet method was used to find the dominant periodicities of magnetic field fluctuations in the magnetosphere of Jupiter. 

The present study examines a machine learning approach to the second category of methods. Specifically, we investigate framing the estimation of high-quality statistics from a dataset with gaps as a supervised learning regression problem. Whereas the mapping from a complete dataset to its statistics is generally in the form of a simple function (e.g., the equation for the mean or standard deviation), we are interested in whether a neural network - the ‘universal approximator’ - can learn a mapping from an \textit{incomplete} dataset to the `clean' statistic that \textit{would} have followed, had the complete dataset been available. This approach is taken because the primary goal is not to accurately reproduce the complete series itself, but rather the statistics calculated from the complete series. Furthermore, we are not aiming to extract the true relationship between input and output, but rather to find an input-output mapping that achieves good performance on unseen gapped datasets; hence the machine learning approach.

As a case study, this technique is applied to time series of the fluctuating interplanetary magnetic field measured by the NASA spacecraft Parker Solar Probe. The interplanetary magnetic field contains many structures and fluctuations, including Alfv\'en wave-like intervals \cite{BelcherJGR1971, TsurutaniGRL2002} and discontinuities \cite{TsurutaniJGR1996, TsurutaniRGP1999}. Although historically the presence of Alfv\'en wave-like structures was thought to imply that the solar wind is not turbulent, there is overwhelming evidence that it is \cite[and references therein]{ColemanApJ68, TuSSR95, MarschLRSP06, BrunoLRSP13, VerscharenLRSP19}. This turbulence is an important ingredient in determining the evolution of the solar wind \cite{UsmanovApJ11, OpherApJ11, sokolov2013, oran2013, BurlagaApJ2018, vanderHolstApJ22}. Hence, to explore the problem of finding good statistics for sparse time series, we work with heliospheric measurements of the turbulent magnetic field. The statistic we attempt to estimate is the \textit{structure function}.

The $n$th order structure function for a time-varying signal $a$ is defined as \cite{BatchelorBook, BiskampBook} 
\begin{equation} S^{(n)}_{a}(\tau) = \av{|\delta a (t,\tau)|^n}\end{equation}
where $a(t)$ is the scalar variable of interest,  $\delta a(t,\tau) = a(t+\tau)-a(t)$ is the increment, $n$ is the order, and $\av{}$ denotes expectation over $t$. For a vector set of time series $\vek{a}(t)$ $=(a_x(t),$ $a_y(t),$ $a_z(t))$, the structure function is defined as 
\begin{equation} S^{(n)}_a(\tau) = \av{|\delta \vek{a}(t,\tau)|^n}\label{eq:strfn}\end{equation}
where $\delta \vek{a}(t,\tau) = \vek{a}(t+\tau)-\vek{a}(t)$. 

\add[DW]{$S^{(n)}_a(\tau)$ gives the $n$-th moment of the probability distribution function of the increments of $a$ at lag $\tau$. Turbulence theory predicts that }the structure functions of various orders \change[DW]{follow}{show} power-law behaviour in the inertial range. \add[DW]{Departures from the expected power-laws across multiple orders are of interest because they represent the presence of intermittent structures in the fluid and the potential need for corrections to the theory of the inertial range \cite{Frisch1995}}\remove[DW]{These statistics, by themselves and in combinations, encode a significant amount of physics, such as the spread of energy across scales and the locality and intermittency of the turbulent structures \cite{BiskampBook, Panchev1971}}. 

In this paper, as a proof-of-concept, we stick to the second-order structure function, i.e., $n=2$ in \change[DW]{Equation (2)}{equation (\ref{eq:strfn})}; this will be referred to as \sfn~for brevity. \add[DW]{This quantity is particularly important due to its close relationship with both the autocorrelation and the energy spectrum \cite{Chhiber2018}. For more detailed discussion of these concepts we refer the reader to} \citeA{MatthaeusJGR82}, \citeA{BiskampBook}, and \citeA{BrunoLRSP13}.

Machine learning algorithms are being widely applied to space weather research, in order to take advantage of the increasing amounts of data and computing power now available \cite{camporeale_19}. These applications have ranged from identifying magnetic reconnection to predicting solar flares \cite{Hu_2020, Bobra_2015}. There are two key challenges with the current application. The first is going from a high-dimensional input space to a high-dimensional output space. Each set of `features' is an entire time series, and each expected output is not a single label or value but rather an array of values: the values of a statistical function. Such a task seems to be unique in the literature, and it is therefore of interest how well a simple network architecture can perform at such a task. The second challenge is finding an appropriate representation of the missing data. Typically, supervised machine learning models require complete data. \add[DW]{For example, the training of a feedforward neural network requires complete inputs in order for the hidden layers to feed forward valid inputs during the forward pass and then update the weights appropriately during the backpropagation step \cite{Brown2003}.} Therefore, it is not immediately obvious how one would use such a model in the second category of techniques\change[DW]{,}{ (i.e.,} without interpolation of the gaps\change[DW]{,}{)} and this remains an open problem in the machine learning community \cite{Sharpe1995, Caiafa2021, Emmanuel2021}. One solution is to use a Cosine Neural Network \cite{gips_2008}. This architecture is able to process and recognise missing data without any prior imputation, thereby addressing the issue of how to represent missing data to a neural network. It does this using `weighted norms', parameters which reduce to zero when the corresponding input feature is missing. This informs the network to ignore these features for that instance. Another approach is to incorporate the probability distribution of missing values into the network inputs \cite{Smieja2018}. 

Alternatively, one could simply use a `placeholder' value that the network would ideally learn to ignore with exposure to the data. Generally, using zero is advised against, due to its tendency to create a discontinuity in the latent space formed by the features, thereby making it harder for the model to generalise \cite{Chollet2021}. Instead, it is recommended to replace the missing values with the average or median value of the dataset. However, for the training dataset used here, zero \textit{is} the average value of each standardized input. On the other hand, imputing an arbitrarily large value, well outside the range of data, could instead teach the network that this value is to be ignored. Given the desire to keep the pipeline as simple as possible for this exploratory study, and that there is no established, universal strategy for dealing with gaps for machine learning models, three different placeholder values were tried: zero, one hundred (i.e., 100$\sigma$ from the mean), and a linear interpolation across the gaps.

To summarise the approach of this study, after the clean, `true' structure functions are calculated for a series of continuous intervals of the time series, the intervals are artificially gapped in several different ways. By creating multiple versions of inputs with gaps in different places and the same expected outputs, we aimed to make the network more robust to missing data. The three types of imputation described above are then applied to the gaps so that the neural network can receive the data. These intervals are provided as the input data to the model, with the original structure functions as the target outputs. The model predictions are then compared with structure functions calculated directly from the gapped intervals, and from gapped intervals with the interpolation techniques applied. In this way we can compare the performance of each technique in approximating the \change[DW]{clean} statistic of the original ungapped interval.

\section{Data Preparation}

\change[DW]{
The data used in this project were taken from the Parker Solar Probe's (PSP) \cite{Fox2016} fluxgate magnetometer (FGM) instrument \cite{Bale2016}. PSP is a spacecraft that was launched in 2018 to study the physics of the inner heliosphere and the origins of the solar wind by flying very close to the Sun (as close as 9.9 solar radii during orbit 22 in 2024). The FGM measures magnetic fields at a native cadence of 256 samples/second. We use data from November 2018, during the first `encounter' (E1) of PSP \cite{BaleNature19, KasperNature19}.  Encounter data are typically at the highest resolution.
}{
Parker Solar Probe is a spacecraft that was launched in 2018 to study the physics of the inner heliosphere and the origins of the solar wind by flying very close to the Sun (as close as 9.9 solar radii during orbit 22 in 2024). Measurements from this spacecraft are of clear scientific interest, and their high resolution provided us with the long continuous series needed for this study. The data used were taken from PSP's fluxgate magnetometer (FGM) instrument \cite{Fox2016, Bale2016}. The FGM measures magnetic fields at a native cadence of 256 samples/second. We use data from November 2018, during and shortly after the first ``encounter'' (E1) of PSP. Encounters are periods centered around each perihelion during which data is typically collected at the highest resolution. For context, during this particular encounter, the Sun was at solar minimum, and PSP recorded measurements at a relatively constant longitude relative to the Sun's rotating surface, travelling as close to the Sun as 35.6 solar radii \cite{BaleNature19, KasperNature19}.
}

Two continuous, gap-free intervals were selected: one from 2018-11-01 to 2018-11-18 and another from 2018-11-21 to 2018-11-30. These intervals contained no missing observations after performing down-sampling, justified in the following section. Combined, these final gap-free time series consisted of 3,100,000 points for each vector component \textit{Bx, By, Bz}, all three of which were used here. These series were then split into 310 vector time series intervals of length 10,000 (duration 125 minutes).

\subsection{Input preparation} 

\begin{figure}[!htb]
    \centering
    \includegraphics[height=\linewidth,
    ]{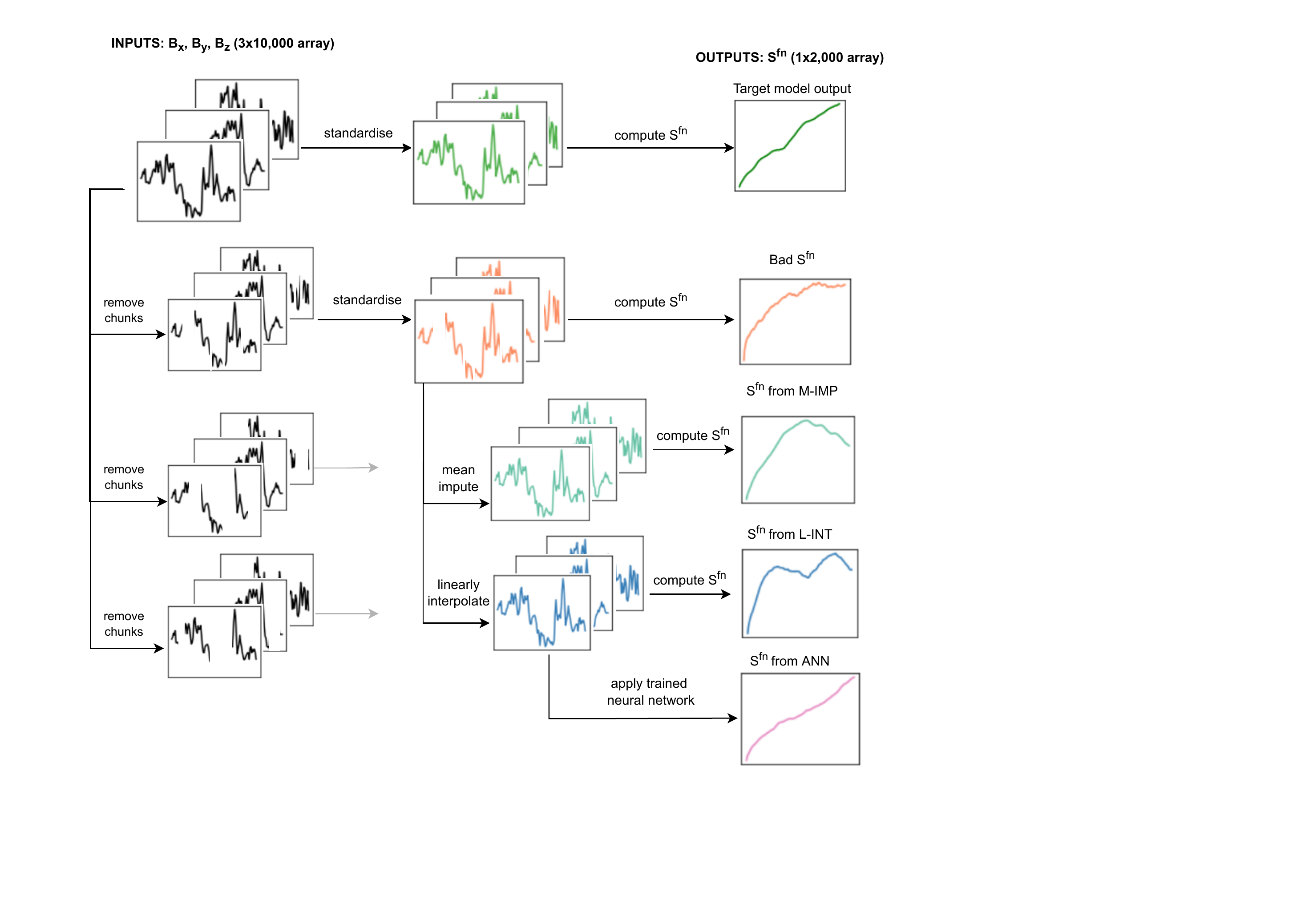}
    \caption{Diagram showing the workflow for adding artificial gaps and producing different structure function estimations from each interval. \add[DW]{(M-IMP = mean imputation, L-INT = linear interpolation.)}}
    \label{fig:sf_workflow}
\end{figure}

The data was pre-processed in order to make it easier for the ANN to process data from different sources and intervals. We started with a set of time series intervals with 100\% of measurements available and followed the data standardisation and augmentation procedure below. Figure \ref{fig:sf_workflow} provides a visual representation of both the data pre-processing and processing workflow.

{\em Data standardisation:} The timescales and magnitudes of interest vary significantly from one system to another. For example, solar wind in the inner heliosphere has magnetic field amplitudes in the $\sim 100 nT$ range and a correlation time of $\approx 600 s$ \cite{ParasharApJS20, ChenApJS20}, whereas the solar wind at 1AU has magnetic field amplitudes in the $\approx 10nT$ range and a correlation time of $\approx 1 hour$ \cite{IsaacsJGR15, JagarlamudiApJ19}. On top of this variability, the time cadences of various instruments differ significantly. In order to train the ANN in a system-agnostic way, we standardised both the x-values (time series) and y-values (the fluctuation amplitudes) using the following methods. The time series were standardised by down-sampling to have 10,000 samples across $\approx 15 t_{corr}$ so that the training series has a sampling rate of $\delta t \sim 1.5\times 10^{-3} T_{corr}$. The amplitudes were standardised by subtracting the mean value \(\mu_a\) from each value of each interval and then dividing by the standard deviation \(\sigma_a\) (Z-score normalisation):

\begin{equation} 
    a(t)_{norm} = \frac{a(t) - \mu_a}{\sigma_a}
\end{equation}

{\em Gapped series preparation and augmentation:} Each set of \change[DW]{clean}{complete} intervals was augmented by duplicating each interval several times and then \change[DW]{made ```dirty''}{corrupted} by taking gaps from each new interval in different ways. By having the same expected output for each of these copies, the aim was to make the ANN indifferent to where and how large the gaps are, as well as giving us more data train on. To this end, 10 copies of the training intervals and 5 copies of the testing and validation intervals were made. Then, a percentage of data was removed from each interval in between 3 and 20 continuous segments. \add[DW]{Real-world data gaps also occur as individual missing values rather than continuous segments. However, large continuous gaps are representative of many bad quality datasets, including OMNI} \cite{Lockwood2019} \add[DW]{and Voyager} \cite{Gallana2016}, \add[DW]{hence the decision to simulate this variety of `missingness'.} 

For the training and validation intervals, between 0 and 50\% of data was removed, whereas for the test intervals, up to 95\% of data was removed. This was done to test the algorithm's performance on missing data \textit{in general}, rather than just the gapped percentages it was trained on. This also allows us to test its performance `in the limit', i.e. right up to only 5\% data remaining, which will help us assess overfitting of the model. \add[DW]{(90\% data loss is also the scale of gaps in Voyager datasets from the outer heliosphere \cite{Gallana2016}.)} For each interval, the exact percentage of data to remove, the number of segments to remove it in, and the location of each segment were chosen randomly. Finally, the resultant \change[DW]{dirty}{gapped} intervals were re-standardised.

{\em ANN input:} The input for the ANN training program was a 3x10,000 array representing all of the three vector components $(Bx,By,Bz)$ across a 10,000 point time series. Three versions of the inputs were created: one with gaps filled with zeroes, another with gaps linearly interpolated, and the third with gaps filled with 100. Note \add[DW]{that} the first two approaches were also used as alternative estimation methods, by simply feeding these input versions into \change[DW]{E}{e}quation (\ref{eq:strfn}), as shown in Figure \ref{fig:sf_workflow}.

{\em ANN output:} The corresponding expected output for the ANN \change[DW]{is the}{was a 1x2,000 array of the values of} second-order \sfn~computed from the corresponding \change[DW]{clean}{complete} time series\change[DW]{The \sfns~were computed}{,} up to a maximum lag of 20\% of the \change[DW]{data size}{series length} ($n_{lag}=2000$)\remove[DW]{\cite{MatthaeusJGR82}}. \add[DW]{(This array can then simply be plotted to show the structure function.)} \remove[DW]{The clean and dirty time series are arranged in random order to create the input matrix, with the appropriate expected outputs.} \note[DW]{Moved to end of following paragraph for better flow}

{\em Benchmarking the results:} 70\% of the intervals were used for training, 20\% for testing, and 10\% for validation of the network: information about each set is summarised in Table \ref{table:model_data}.  The second-order \sfns~computed by the ANN are then compared to the \sfns~computed three ways: i) ignoring the gaps, ii) mean imputing the gaps, and iii) linearly interpolating the gaps, as shown in Figure \ref{fig:sf_workflow}. For computation i), this is achieved by simply ignoring the values for which no lagged increment is available. \add[DW]{Finally, the input-output pairs are arranged in random order before training.}

\begin{table}[h]
\centering
\begin{tabular}{|p{0.08\linewidth}|c|p{0.14\linewidth}|p{0.14\linewidth}|p{0.16\linewidth}|p{0.14\linewidth}|c||} 
 \hline
 Data source & Purpose & Input lengths & Output lengths ($n_{lag}$) & \% of each input removed &  No. of instances\\ 
\hline
  PSP & Training set & 30000 & 2000 & 0-50 & 2170\\ 
     & Validation set & 30000 & 2000 & 0-50 & 150\\
     & Test set & 30000 & 2000 & 0-95 & 315\\
     \hline
\end{tabular}
\caption{Dimensions of data used to build and evaluate the neural network model. `No. of instances' refers to the count of intervals in the set after duplication of the original unique intervals.}
\label{table:model_data}
\end{table}

\section{Model training and validation}

Using a feedforward neural network, a multi-output regression model was built in Python using the Tensorflow package \cite{tensorflow2015-whitepaper}. The workflow to ensure a good model fit was the following:

\begin{enumerate}
    \item Train the model until the early-stopping criterion is reached (see below)
    \item Evaluate the model on the validation set, checking for overfitting and underfitting by visual inspection of the predictions
    \item Adjust the model hyperparameters, such as the number of hidden layers and the number of nodes in each hidden layer
    \item Repeat 2-3 until a good fit is achieved
\end{enumerate}

The loss function used to calculate the error for this network was the mean squared error, or MSE. Due to each output being a vector, the overall error for one epoch of the network $MSE_{overall}$ is calculated as the MSE for a single instance $MSE_i$, averaged over all the instances. (One epoch is one iteration through every instance in the training set.) 


\begin{eqnarray} \label{eq:MSE}
    MSE_{i} & = &\frac{1}{n_{lag}}{\sum_{j=1}^{n_{lag}} (S^{fn}_{ij, pred} - S^{fn}_{ij, true})^2} 
\end{eqnarray}

where $S^{fn}_{ij, pred}$ is the predicted value of the second-order \sfn~for the $i^{th}$ interval at lag $j$, $S^{fn}_{ij, true}$ is the corresponding `ground truth' value, and $n_{lag}$ is the number of lags for which the \sfn~has been computed.
$MSE_{overall}$ is minimised through the process of backpropagation of error via gradient descent, and each weight and bias is adjusted according to the learning rate and the weight's contribution to the overall error, calculated using partial derivatives. 

For each epoch of training, the \remove[DW]{training loss and the} validation loss \change[DW]{were}{was} used to check whether the model is still improving. (A sustained increase in the validation loss indicates that the model is beginning to overfit.) Accordingly, training was stopped when the validation loss was reduced by no more than 0.01 over 10 epochs\change[DW]{, which we call the EarlyStopping criterion.}{; this is the early-stopping criterion.}

The \sfn~values decrease by a few orders of magnitude going from large to small lags. This could potentially bias the MSE on large-scale predictions. Hence, the MSE was not used in isolation to evaluate the final model on the test set. It was complemented with the mean absolute percentage error (MAPE) to quantify the model's performance. 

\begin{eqnarray} \label{eq:MAPE}
    MAPE_{i} & = & \frac{1}{n_{lag}} \sum_{j=1}^{n_{lag}} 
            \left|\frac{S^{fn}_{ij, true} - S^{fn}_{ij, pred}}
            {S^{fn}_{ij, true}}\right|
\end{eqnarray}

where symbols have the same meaning as in the expression for $MSE_i$. This \change[DW]{loss function}{metric} is easier to interpret than the MSE in terms of relative error and is scale-independent. It is also more stringent than MSE because it evaluates the method's performance in predicting both the large-scale and small-scale values. \add[DW]{MAPE was also trialled as a loss function for training the network; however, this led to inferior performance and overfitting compared with MSE. Therefore, we only present results for the model trained using MSE.}

\subsection{Optimising the network: selection of hyperparameters}

Part of the challenge of supervised learning is finding the optimal hyperparameters (`architecture' in the case of the dimensions of a neural network) so as to get the best performance on unseen data. A common approach is to use a random grid search, which involves randomly selecting a subset of combinations of hyperparameters from a search space, and choosing the combination that produces the minimum validation loss. This method was tried for each of the input versions. The number of hidden layers, number of nodes in each hidden layer, learning rate, and presence of a drop-out layer were varied, whereas the optimizer (Adam) and activation functions (ReLU for each hidden layer and linear for the output layer \cite{Brownlee2022}) were kept the same. However, it was found that this traditional approach of optimisation was not suitable for this unusual prediction task. Aiming for the smallest loss (average MSE) on the validation set led to a `regression to the mean' style of overfitting, where all predicted curves were the same. In other words, the resultant network was only paying attention to the biases of the final layer, rather than the actual inputs. \add[DW]{(}The networks that produced these lowest losses (between 0.68 and 1.2 average MSE) had optimal hyperparameters of between 9 and 17 layers, depending on the input version used.\add[DW]{)} This result called for a more careful optimisation procedure. Specifically, the best network architecture was selected by manually increasing the size of the network and visually inspecting the predictions of each network on a sample of validation instances, rather than relying on the validation loss as an indicator of model performance.

\begin{figure}[!htb]
    \includegraphics[width=1.\linewidth]{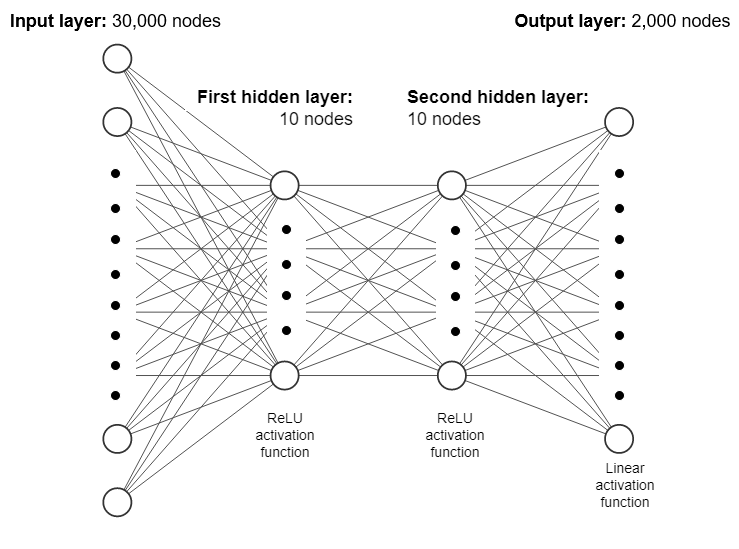}
    \caption{Schematic diagram of the final ANN architecture. The inputs consist of 3 x 10,000 stacked vector components, and the outputs consist of 1 x 2,000 second-order structure functions. The number of hidden layers and nodes were chosen by iteratively increasing the size of each hyperparameter and visually inspecting a sample of validation predictions} made by the trained network for each architecture.
    \label{fig:schematic_ann}
\end{figure}

After running several different iterations it was found that 10 or more layers always lead to regression to the mean, where virtually the same curve was predicted for every input interval. The best network configuration from those that were tried - i.e., that which produced smooth curves with shapes that at least partly matched the different shapes of the expected outputs -  was one with 2 hidden layers, each with 10 nodes; a learning rate of 0.001; no drop-out layer; and trained for 500 epochs. A schematic diagram of this configuration is shown in Figure \ref{fig:schematic_ann}.  We note that the input and the output layers require the largest number of parameters, 300,010 and 22,000, respectively in this case. Any additional hidden layers having order 10 neurons contribute only order 100 parameters to the list of trainable parameters. \add[DW]{Finally,} the input version that produced the best predictions was the linearly interpolated dataset.

\section{Results} 

\begin{figure}[!htb]
    \includegraphics[width=1.\linewidth]{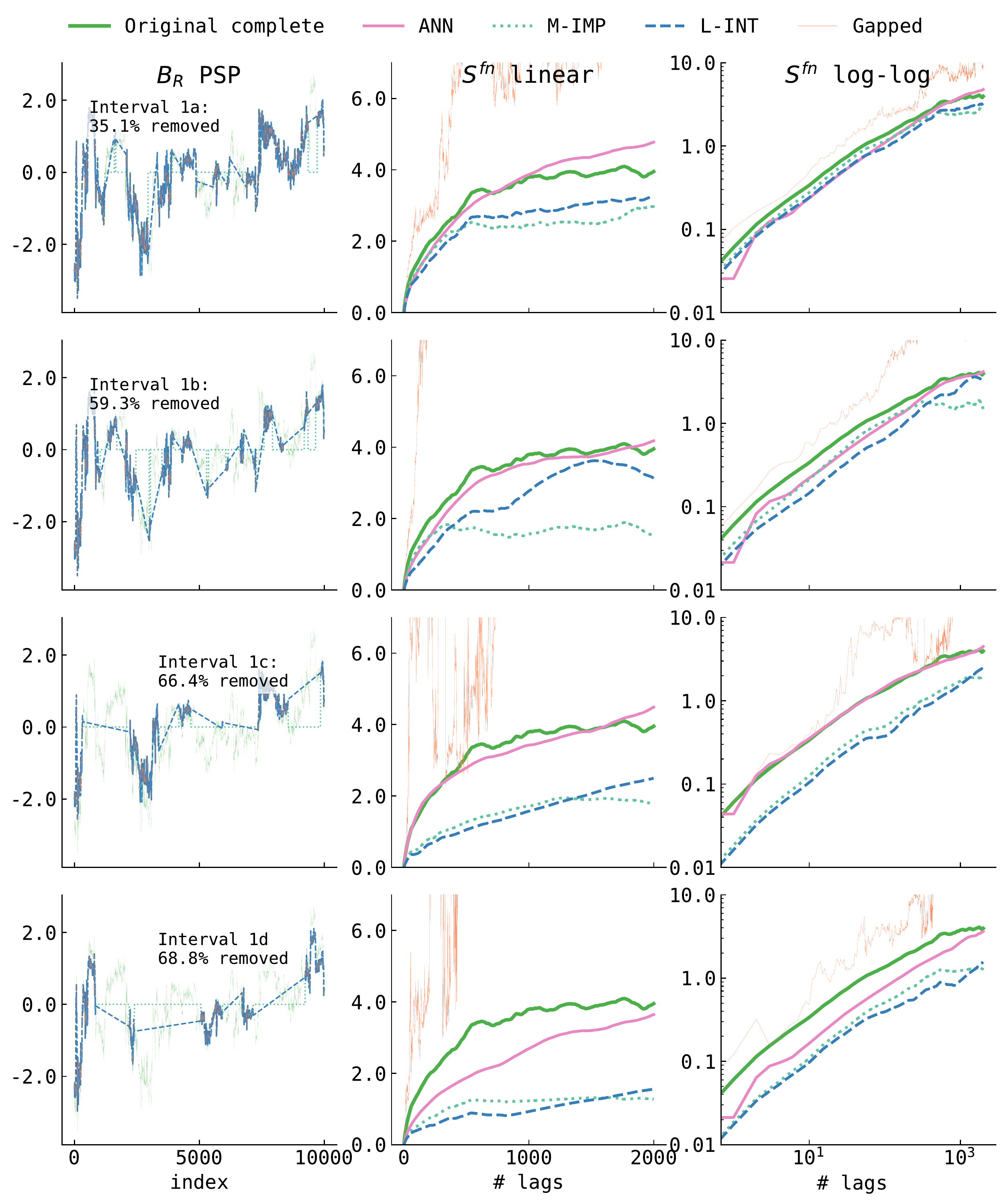}
    \caption{Results of the four different \sfn~approximation methods for a standardised PSP interval that has been gapped in four different ways. (For simplicity, only one of three vector components for each input interval has been shown, but this still illustrates the number and size of the gaps, which were consistent between components.) Note that the log-log plot emphasises differences between the curves at small lags (i.e., small spatial scales)\add[DW]{, and is the standard way of presenting structure functions, in order to facilitate comparison with theoretical power-laws.}}
    \label{fig:psp_case_studies}
\end{figure}

The favoured network architecture and input version were then used to produce predictions on the test set and ultimately evaluate the model's performance. Its approximation ability was compared with the alternative interpolation methods discussed previously, as well as naive calculation of the structure functions directly from the gapped intervals (see Figure \ref{fig:sf_workflow}). We start with a case study in Figure \ref{fig:psp_case_studies} of four versions of the same original \change[DW]{clean}{complete} interval, each with increasing data loss. This is one example of how the network responds to the input data; an overall evaluation of its ability follows.

\remove[DW]{For input interval 1a, with 35\% data missing, the ANN does not accurately reproduce the flattening of the true structure function above about lag 500. This contrasts with M-IMP and L-INT, which do show the correct shape, albeit shifted downwards. The prediction for 1b, at 59\% data loss, gets closer to the true shape, and here we see the other methods noticeably degrade in their estimation. The log-log plot shows how the ANN particularly has better predictions at larger scales (higher lags). At 66\% data loss, the ANN shows excellent performance, relative to other methods. In this case it predicts the true values well across the range of scales - this is particularly evident in the log-log plot. Finally, at 69\% data loss, the performance of the ANN poorly follows the shape of the true curve, though does still get closer than other methods, especially at large scales. Given that the \sfn~decreases by a few orders of magnitude from large to small lags, small lags contribute very little to MSE. Hence, it is expected that such predictions could produce better MSE performance for the ANN when large amounts of data are missing. In contrast, L-INT and M-IMP show a predictable consistent underestimation, due to the decreased variability in magnetic field increments caused by the smoothing effect of the imputation. This effect worsens with increasing gap percentage. This foreshadows the ultimate conclusion on the utility of the neural network, which seems to perform best, relative to other methods, when dealing with large data loss.
}{
\add[DW]{\textit{Performance of ANN on case study:}} For input interval 1a, with 35\% data missing, the ANN does not accurately reproduce the flattening of the true structure function above about lag 500. The prediction  interval 1b, at 59\% data loss, gets closer to the true shape. The log-log plot shows how the ANN particularly has better predictions at larger scales (higher lags). At 66\% data loss (interval 1c), the ANN shows excellent performance, relative to other methods. In this case it predicts the true values well across the range of scales - this is particularly evident in the log-log plot. Finally, at 69\% data loss (interval 1d), the performance of the ANN poorly follows the shape of the true curve, though does still get closer than other methods, especially at large scales. Given that the \sfn~decreases by a few orders of magnitude from large to small lags, small lags contribute very little to MSE. \change[DW]{Hence, it is expected that such predictions could produce better MSE performance for the ANN when large amounts of data are missing.}{Hence, it is expected that, given this loss function, values at larger lags would be better predicted by the neural network.} This foreshadows the ultimate conclusion on the utility of the neural network, which seems to perform best, relative to other methods, when dealing with \change[DW]{large data loss}{large-scale values at high data loss}.

\add[DW]{\textit{Performance of simple imputation on case study:}} At 35\% data loss, M-IMP and L-INT, do show the correct shape, albeit shifted downwards. At 59\% missing, these methods noticeably degrade in their estimation of this structure function, and at 66\% missing they are clearly inferior to the ANN. With just slightly more missing data (interval 1d), M-IMP and L-INT show little change, retaining their severe under-estimation of the true curve. In contrast to ANN, L-INT and M-IMP show a predictable consistent underestimation, due to the decreased variability in magnetic field increments caused by the smoothing effect of the imputation. This effect worsens with increasing gap percentage.
}

\begin{figure}[!htb]
    \centering
    \includegraphics[width=1\linewidth]{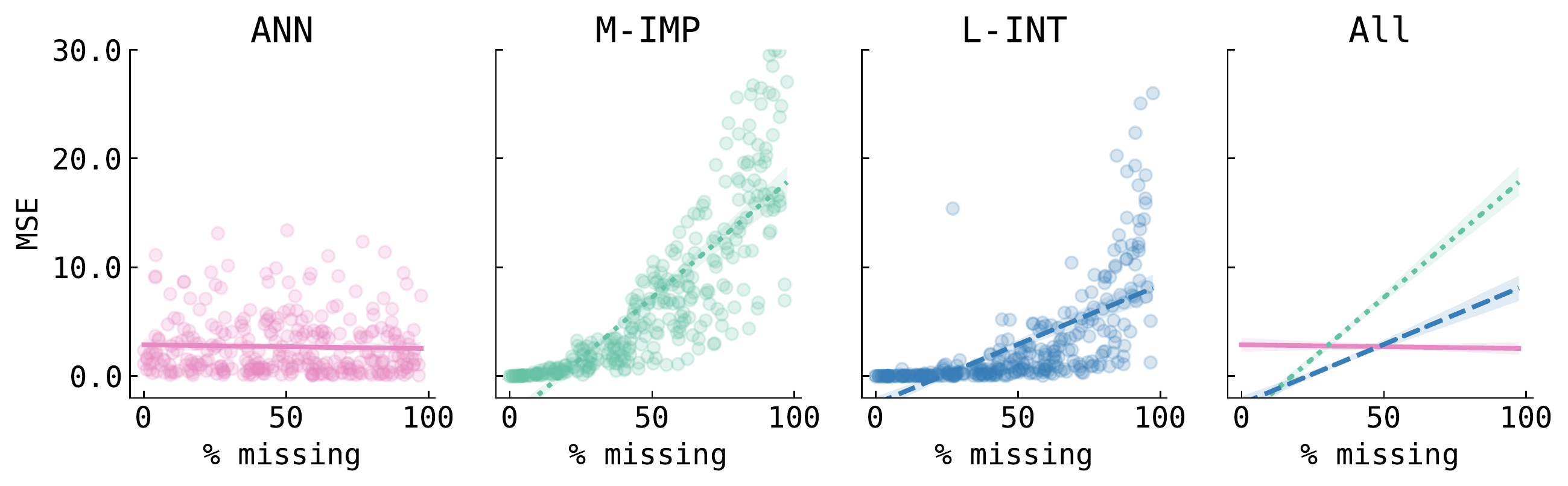} 
    \caption{Scatter plot of MSE against proportion of data removed for the PSP test intervals with overlaid linear regression lines and confidence regions from other panels. The line for the GAPS method (no imputation) is not shown here as it quickly disappears from the plotting area, and Table \ref{table:results} shows it is clearly inferior to the other methods.}
    \label{fig:psp_mse}
\end{figure}

\begin{figure}[!htb]
    \centering
    \includegraphics[width=1\linewidth]{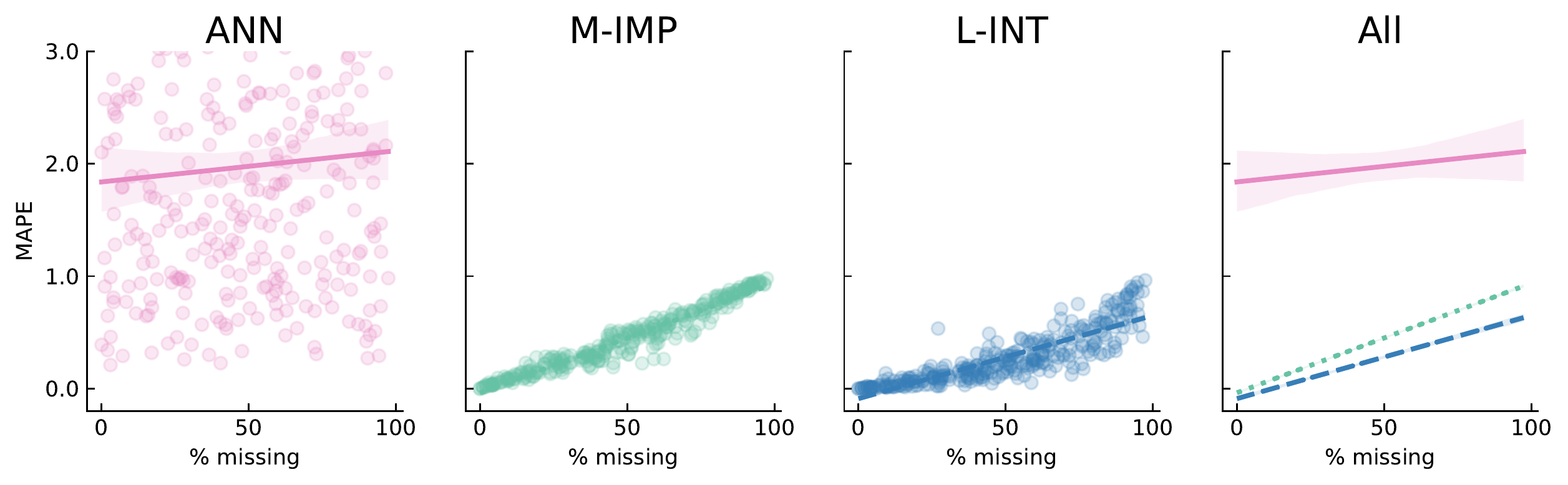} 
    \caption{Scatter plot of MAPE against proportion of data removed for the PSP test intervals with overlaid linear regression lines and confidence regions. The line for the GAPS method (no imputation) is not shown here as it quickly disappears from the plotting area shows it is clearly inferior to the other methods.}
    \label{fig:psp_mape}
\end{figure}

\add[DW]{\textit{Overall statistical performance: }}The overall performance of each method, including the \sfns~calculated from the un-filled gapped intervals, was evaluated using the following measures:
\begin{itemize}
\item Average MSE (\(MSE_{overall} = \frac{1}{n}{\sum_{i=1}^{n} MSE_i}\)) across all expected-observed \sfn~pairs (recall this was the loss function used to train the neural network - see \change[DW]{Equation \ref{eq:MSE}}{equation (\ref{eq:MSE}))}). 
\item Average MAPE (\(MAPE_{overall} = \frac{1}{n}\sum_{i=1}^{n} MAPE_i\)) across all expected-observed \sfn~pairs. Both $MSE_{overall}$ and $MAPE_{overall}$ are given in Table \ref{table:results}.

\item Scatterplots and corresponding linear regression lines of MSE and MAPE against \% data missing of individual test intervals. This shows us the how each method performs with increasing data loss shown in Figures \ref{fig:psp_mse} and \ref{fig:psp_mape}.
\end{itemize}

\begin{table}[h]
\centering
\begin{tabular}{|c|c|c|c|c|c|} 
 \hline
 \textbf{Spacecraft} & \textbf{Performance} & \multicolumn{3}{|c|}{\textbf{\sfn calculated from}} & \textbf{\sfn estimated using} \\
 & \textbf{measure} & \textbf{GAPS} & \textbf{M-IMP} & \textbf{L-INT} & \textbf{ANN} \\
 \hline
 \hline
 PSP & MSE & 16536.91 & 7.02 & 2.82 & \textbf{2.71} \\ 
     & MAPE & 5.13 & 0.44 & \textbf{0.27} & 1.97 \\
     \hline
\end{tabular}
\caption{Calculated performance measures for each \sfn~approximation method. \textbf{Bolded figures are the lowest of each row.} GAPS: Gapped interval with no imputation. M-IMP: Gapped interval with mean (0) imputation. L-INT: Gapped interval with linear interpolation of gaps. ANN: Artificial neural network model.}
\label{table:results}
\end{table}

The overall results shown in Table \ref{table:results}, suggest that, for overall average performance, the GAPS (calculation from gapped series) method is the worst \sfn~approximation method across the board with $MSE_{overall}\approx17,000$ and $MAPE_{overall}\approx5.1$. The relative performance of the other methods differ depending on the evaluation metric used. This occurs because MAPE is lower for predictions that better estimate the true curve as a proportion at each lag, whereas MSE is lower for predictions that simply have the smaller average squared distance from the true curve. Because of this difference, \change[DW]{the ANN has the best overall performance according to MSE}{according to MSE, the ANN has the best overall performance}, followed by L-INT\change[DW]{, but L-INT has the best performance according to MAPE, followed by M-IMP.}{. According to MAPE, L-INT has the best performance, followed by M-IMP, and the ANN is only the third best.}

However, these overall measures do not take into account the variation in their performance as a function of the degree of sparsity. Hence, we take a statistical approach to quantify the performance with increasing sparsity. Figure \ref{fig:psp_mse} shows the scatter plots of MSE versus percentage of missing data for each set of test intervals for each method, overlain with linear regression lines of best fit. As seen in the middle two panels, the MSE tends to increase with increasing sparsity for M-IMP and L-INT. This is to be expected: as the amount of data missing increases, the \sfn~estimation gets worse for these methods. However, there is distinct funnelling on the plots, representing heteroskedasticity or unequal variance in MSE for different proportions missing. For low \% missing values, around less than 20\% missing, there is a very small range of MSE values for the imputation methods. This means that the accuracy of the \sfn~estimations do not vary much for small percentages of missing data. On the other hand, as the amount of missing data increases, not only does the average error increase, but also the variation in error. This shows that intervals with large amounts of data missing can, in some cases, produce \sfns~quite similar to the true \sfns, if imputation is performed. This is especially true for intervals that have been linearly interpolated - we can see in the L-INT scatterplot that there are intervals with up to 90\% missing that have very low MSE values. This is due to the importance of not only the size of the gaps, but where the gaps are in the interval: if the removed data does not significantly depart from the overall trend, linear interpolation will result in a \sfn~not very different from the expected curve. On the other hand, M-IMP does not show similarly low MSE values beyond about 45\% missing, though there is still distinctly increasing variance.

In stark contrast to the M-IMP and L-INT methods, the MSE of the ANN model predictions are largely indifferent to the proportion of data missing. There is a constant band of MSE values across this scatterplot, and the Pearson correlation coefficient, a measure of the linear association between two variables, is very close to 0 for this method (-0.06).

As a way of establishing the comparative usefulness of each method, linear regression lines were fit to the data. Although fitting a linear model is inappropriate for this data due to the unequal variance present in the M-IMP and L-INT methods, it still provides a useful indicator of the typical values of MSE for different proportions missing. \change[DW]{What really distinguishes these methods, as shown in Figure \ref{fig:psp_mse}, are the slopes of the regression lines.}{The final panel in Figure \ref{fig:psp_mse} shows a clear distinction in the slopes of the regression lines.} MSE increases the fastest for the M-IMP method, and this association is also that with the highest correlation between MSE and \% missing (0.87). \change[DW]{Next is}{The next largest slope is that of} the L-INT method, and this association has the next highest correlation (0.70). Finally, the ANN has a very flat slope, with an aforementioned correlation of close to 0. This result shows that the ability of the neural network to approximate the true \sfn~is much less affected by increasing amounts of missing data than the other two methods. However, this does not make it the best \sfn~approximation method for any bad dataset. What we can see in Figure \ref{fig:psp_mse} is that up to about 50\% missing data, the simple imputation methods have lower typical values of MSE than the neural network. At values greater than 50\%, the neural network, on average, produces the best estimations of the three approaches, according to the MSE metric. This, however, could be a result of the dominance of large lag values of the \sfn~controlling the MSE. 

To quantify the performance of a method to predict not only the large lag value but also the small lag values, we use the MAPE. The MAPE measure, shown in Figure \ref{fig:psp_mape}, shows much less heteroskedasticity for the L-INT and M-IMP methods, with M-IMP in particular showing a much stronger linear relationship between \% missing and MAPE, producing a correlation of 0.98. The ANN method has a small positive slope and a relatively constant band of scatter, with correlation of 0.04. \change[DW]{Overlaying the regression lines in}{In the final panel of} Figure \ref{fig:psp_mape}, we see that the neural network linear regression line remains above that of both other methods for all gap percentages. \remove[DW]{Based on these findings, it seems reasonable to use L-INT for data gaps as large as $\sim 20\%$.}\note[DW]{Moved to end of following paragraph for better flow}

Overall, we find that the ANN is largely indifferent to the proportion of data missing in its approximations of the true \sfn. However, this means that the worst predictions for inputs with little or no data missing are as bad as those for inputs with at least 90\% data missing. This means that while this small neural network \textit{can} produce good predictions at lower data loss, it should not be relied upon to do so, unlike the simpler imputation methods. \add[DW]{Based on these findings, it seems reasonable for this task to use linear interpolation for data gaps as large as $\sim 20\%$.}

\section{Summary and Conclusions}\label{section:summary}
Gaps are a common problem in almost all fields that deal with time series. The field is mature with many ways of filling the gaps, including mean-imputation, linear interpolation, maximum likelihood estimation of ARIMA models \cite{Velicer2005}, singular spectrum analysis \cite{Schoellhamer2001}, and artificial neural networks (ANNs) \cite{Comerford2015}. Our interest here is not in prediction, but in gleaning the best estimates of scale-domain statistics of a system without having to reconstruct the time series. To this end, we studied the potential of simple feedforward artificial neural networks to predict turbulent statistics of solar wind magnetic field measurements. In space plasma physics, an accurate description of the inertial range of the structure function is desirable. This is particularly important to estimate not only the slope of the structure function in this range, but also to estimate the inner and outer scales of turbulence.

Starting with \change[DW]{complete}{complete} time series with 100\% coverage, we created \change[DW]{``bad''}{gapped} time series for which the second-order structure functions were estimated in four ways: i) direct computation ignoring gaps, ii) mean imputation of the gaps, iii) linear interpolation across the gap, and iv) a trained ANN.

ANNs do not seem to be the panacea that one might naively hope for in such a situation. Our results showed a narrow range of conditions in which the supervised learning model could be useful. As reflected by the error functions of MSE and MAPE, the ANN seems to somewhat learn to estimate the large-scale values of the structure function. This is not very surprising as the large lag \sfn~values approach the mean-squared value of the fluctuation amplitudes. Given the trend in error with increasing data loss, the ANN is more useful for large portions of missing data, but over the entire range of data loss it tends to perform worse than simpler methods. 

However, it is worth noting that with only 20 neurons (and about 322,000 trainable parameters), ANN performs comparably to L-INT or M-IMP methods, and with MSE as the cost function it even outperforms these two methods for large gaps in the data. However, when using MAPE to evaluate the network we find the ANN consistently falling short compared with other methods, indicating its lack of ability to predict the structure function accurately at different scales, according to this \textit{proportional} measure of accuracy. \add[DW]{In addition, the model was only evaluated on data from the same system as it was trained on (Encounter 1 of PSP). While we did take the step of re-sampling the input data based on the correlation time of PSP data, we cannot yet make claims to the model's ability to generalise to other space weather domains, for example, the near-Earth environment. Future iterations could train the model on data from a range of spacecraft and solar conditions to attempt to improve its performance on diverse unseen data.}

Our results indicate that to achieve a reasonable description of turbulent statistics for gapped time series, one needs to go beyond simple-minded feedforward ANNs. \remove[DW]{Possible improvements to ANNs could include grey-box modeling with turbulence physics incorporated into the input, and more advanced architectures such as LSTM networks or autoencoders.}\add[DW]{Possible improvements to ANNs could include more advanced architectures such as LSTM networks or autoencoders, and grey box modeling with turbulence physics incorporated into the input. This study has taken a reasonably physics-agnostic approach to training the model, but a grey box model that incorporates additional physical parameters and constraints could potentially improve its performance. Such an approach could also be applied to the adjacent problem of identifying waves in the solar wind. The presence and nature of wave-like helical structures are measured by the magnetic helicity, and this helps to identify the nature of turbulence  \cite{TsurutaniJGR2018}. Magnetic helicity spectra have been used by} \citeA{Podesta_2011}, \citeA{He_2011}, and \citeA{Klein_2014} \add[DW]{to identify the presence of ion cyclotron waves and kinetic Alfv\'en waves in the solar wind. This represents an ideal problem for grey box modeling, where the helicity signatures guide the machine learning algorithms to quantitatively describe solar wind plasma properties. Another example of helicity, the total unsigned current helicity, has been shown to be an important feature in multiple solar flare forecasting models, including a CNN model \cite{Yi2021}. CNNs have also been used to distinguish between sub-Alfvenic and super-Alfvenic MHD turbulence using snapshots of simulations \cite{peek2019}.}

It may also be the case that even the performance of this simple feedforward ANN structure could be improved through better optimisation of the model weights, biases, and hyperparameters (in particular, the number of hidden layers and nodes). To this end, a reliable method of avoiding over-fitting when trying to predict the shape of a curve, rather than a scalar output, is an important issue to address. The conventionally used loss function ($MSE_{overall}$) is a very limited measure because it can cause the neural network to regress to the mean and essentially ignore the input values. This will be explored in follow-up studies, along with other ways of representing the missing data when feeding it into the network. On this note, our results also suggest that not only does linear interpolation remain a reliable method of dealing \add[DW]{with} relatively small data gaps, but it is also a useful and simple way of preparing a gapped series for input into a machine learning model. Imputing 100$\sigma$ and 0 in place of the gaps failed in this regard, likely due to the network's failure to both detect and discount the effect of these `placeholder' values during the processing of the input vector. This is a significant challenge for any mapping technique that builds in local linearity, as neural nets do.

The recent advances in applying machine learning to space physics provide exciting avenues for exploration and discovery. The literature is expanding rapidly and many interesting applications remain, such as identifying discontinuities \cite{TsurutaniRGP1999} and helical structures \cite{TsurutaniJGR2018} in the solar wind. Progress has already been made in classifying states of the wind \cite{Camporeale2017}, as well as improving the resolution of solar images through deep learning \cite{Kim2019} and extracting features from these images for flare prediction \cite{Jiao2020}. The playground is big and largely unexplored, and could lead to breakthroughs in understanding and predicting the space weather environment. In order to do this, we need a technique for dealing with significant data sparsity in time series and the attendant degradation of models and statistics. This study has examined a machine learning approach to this issue, and in doing so aimed to make the literature less `sparse'.

\section{Author Contributions}
TNP came up with the project idea, DW performed the analysis and created the figures, RR, MF, and RKS provided guidance on ANNs. All authors discussed the results and contributed to manuscript writing.

\section{Data Availability Statement}

The raw data can be downloaded from the Space Physics Data Facility website: \url{https://spdf.gsfc.nasa.gov/pub/data/psp/fields/l2/mag_rtn/}. The processed data and software are available on Zenodo and Github: \url{https://doi.org/10.5281/zenodo.6960975}. This work used computational and storage services associated with the Rāpoi cluster provided by Victoria University of Wellington.

\acknowledgments

We would like to acknowledge the PSP instrument teams for high quality measurements in the inner heliosphere and the Space Physics Data Facility (SPDF) at the Goddard Space Flight Center for providing access to the data used for this study. This research was seeded by funding from a Summer Research Scholarship provided by Victoria University of Wellington. The work of RKS is partially supported by SERB, DST, Government of India through the project EMR/2017/002778.


%
%

\bibliography{WrenchANN}

\begin{thebibliography}{}

\bibitem [\protect \citeauthoryear {%
Abadi%
\ \protect \BOthers {.}}{%
Abadi%
\ \protect \BOthers {.}}{%
{\protect \APACyear {2015}}%
}]{%
tensorflow2015-whitepaper}
\APACinsertmetastar {%
tensorflow2015-whitepaper}%
\begin{APACrefauthors}%
Abadi, M.%
, Agarwal, A.%
, Barham, P.%
, Brevdo, E.%
, Chen, Z.%
, Citro, C.%
\BDBL {}Zheng, X.%
\end{APACrefauthors}%
\unskip\
\newblock
\APACrefYearMonthDay{2015}{}{}.
\newblock
\APACrefbtitle {{TensorFlow}: Large-Scale Machine Learning on Heterogeneous
  Systems.} {{TensorFlow}: Large-scale machine learning on heterogeneous
  systems.}
\newblock
\begin{APACrefURL} \url{https://www.tensorflow.org/} \end{APACrefURL}
\PrintBackRefs{\CurrentBib}

\bibitem [\protect \citeauthoryear {%
Bale%
\ \protect \BOthers {.}}{%
Bale%
\ \protect \BOthers {.}}{%
{\protect \APACyear {2019}}%
}]{%
BaleNature19}
\APACinsertmetastar {%
BaleNature19}%
\begin{APACrefauthors}%
Bale, S.%
, Badman, S.%
, Bonnell, J.%
, Bowen, T.%
, Burgess, D.%
, Case, A.%
\BDBL {}others%
\end{APACrefauthors}%
\unskip\
\newblock
\APACrefYearMonthDay{2019}{}{}.
\newblock
{\BBOQ}\APACrefatitle {Highly structured slow solar wind emerging from an
  equatorial coronal hole} {Highly structured slow solar wind emerging from an
  equatorial coronal hole}.{\BBCQ}
\newblock
\APACjournalVolNumPages{Nature}{576}{7786}{237--242}.
\PrintBackRefs{\CurrentBib}

\bibitem [\protect \citeauthoryear {%
Bale%
\ \protect \BOthers {.}}{%
Bale%
\ \protect \BOthers {.}}{%
{\protect \APACyear {2016}}%
}]{%
Bale2016}
\APACinsertmetastar {%
Bale2016}%
\begin{APACrefauthors}%
Bale, S.%
, Goetz, K.%
, Harvey, P.%
, Turin, P.%
, Bonnell, J.%
, Dudok~de Wit, T.%
\BDBL {}others%
\end{APACrefauthors}%
\unskip\
\newblock
\APACrefYearMonthDay{2016}{}{}.
\newblock
{\BBOQ}\APACrefatitle {The {FIELDS} instrument suite for {Solar Probe Plus}}
  {The {FIELDS} instrument suite for {Solar Probe Plus}}.{\BBCQ}
\newblock
\APACjournalVolNumPages{Space science reviews}{204}{1}{49--82}.
\PrintBackRefs{\CurrentBib}

\bibitem [\protect \citeauthoryear {%
Batchelor%
}{%
Batchelor%
}{%
{\protect \APACyear {1953}}%
}]{%
BatchelorBook}
\APACinsertmetastar {%
BatchelorBook}%
\begin{APACrefauthors}%
Batchelor, G\BPBI K.%
\end{APACrefauthors}%
\unskip\
\newblock
\APACrefYear{1953}.
\newblock
\APACrefbtitle {The Theory of Homogeneous Turbulence} {The theory of
  homogeneous turbulence}.
\newblock
\APACaddressPublisher{New York}{Cambridge University Press}.
\PrintBackRefs{\CurrentBib}

\bibitem [\protect \citeauthoryear {%
Bavassano%
, Dobrowolny%
, Mariani%
\BCBL {}\ \BBA {} Ness%
}{%
Bavassano%
\ \protect \BOthers {.}}{%
{\protect \APACyear {1982}}%
}]{%
bavassano}
\APACinsertmetastar {%
bavassano}%
\begin{APACrefauthors}%
Bavassano, B.%
, Dobrowolny, M.%
, Mariani, F.%
\BCBL {}\ \BBA {} Ness, N\BPBI F.%
\end{APACrefauthors}%
\unskip\
\newblock
\APACrefYearMonthDay{1982}{}{}.
\newblock
{\BBOQ}\APACrefatitle {Radial Evolution of Power Spectra of Interplanetary
  {Aflvenic} Turbulence} {Radial evolution of power spectra of interplanetary
  {Aflvenic} turbulence}.{\BBCQ}
\newblock
\APACjournalVolNumPages{Journal of Geophysical Research}{87}{A5}{3617-3622}.
\PrintBackRefs{\CurrentBib}

\bibitem [\protect \citeauthoryear {%
Belcher%
\ \BBA {} Davis~Jr%
}{%
Belcher%
\ \BBA {} Davis~Jr%
}{%
{\protect \APACyear {1971}}%
}]{%
BelcherJGR1971}
\APACinsertmetastar {%
BelcherJGR1971}%
\begin{APACrefauthors}%
Belcher, J.%
\BCBT {}\ \BBA {} Davis~Jr, L.%
\end{APACrefauthors}%
\unskip\
\newblock
\APACrefYearMonthDay{1971}{}{}.
\newblock
{\BBOQ}\APACrefatitle {Large-amplitude Alfv{\'e}n waves in the interplanetary
  medium, 2} {Large-amplitude alfv{\'e}n waves in the interplanetary medium,
  2}.{\BBCQ}
\newblock
\APACjournalVolNumPages{Journal of Geophysical Research}{76}{16}{3534--3563}.
\PrintBackRefs{\CurrentBib}

\bibitem [\protect \citeauthoryear {%
Biskamp%
}{%
Biskamp%
}{%
{\protect \APACyear {2003}}%
}]{%
BiskampBook}
\APACinsertmetastar {%
BiskampBook}%
\begin{APACrefauthors}%
Biskamp, D.%
\end{APACrefauthors}%
\unskip\
\newblock
\APACrefYear{2003}.
\newblock
\APACrefbtitle {Magnetohydrodynamic turbulence} {Magnetohydrodynamic
  turbulence}.
\newblock
\APACaddressPublisher{}{Cambridge University Press}.
\PrintBackRefs{\CurrentBib}

\bibitem [\protect \citeauthoryear {%
Bobra%
\ \BBA {} Couvidat%
}{%
Bobra%
\ \BBA {} Couvidat%
}{%
{\protect \APACyear {2015}}%
}]{%
Bobra_2015}
\APACinsertmetastar {%
Bobra_2015}%
\begin{APACrefauthors}%
Bobra, M\BPBI G.%
\BCBT {}\ \BBA {} Couvidat, S.%
\end{APACrefauthors}%
\unskip\
\newblock
\APACrefYearMonthDay{2015}{}{}.
\newblock
{\BBOQ}\APACrefatitle {SOLAR FLARE PREDICTION USING {SDO/HMI} VECTOR MAGNETIC
  FIELD DATA WITH A MACHINE-LEARNING ALGORITHM} {Solar flare prediction using
  {SDO/HMI} vector magnetic field data with a machine-learning
  algorithm}.{\BBCQ}
\newblock
\APACjournalVolNumPages{The Astrophysical Journal}{798}{2}{135}.
\newblock
\begin{APACrefURL} \url{https://doi.org/10.1088/0004-637x/798/2/135}
  \end{APACrefURL}
\newblock
\begin{APACrefDOI} \doi{10.1088/0004-637x/798/2/135} \end{APACrefDOI}
\PrintBackRefs{\CurrentBib}

\bibitem [\protect \citeauthoryear {%
Broersen%
}{%
Broersen%
}{%
{\protect \APACyear {2006}}%
}]{%
Broersen2006}
\APACinsertmetastar {%
Broersen2006}%
\begin{APACrefauthors}%
Broersen, P\BPBI M.%
\end{APACrefauthors}%
\unskip\
\newblock
\APACrefYearMonthDay{2006}{}{}.
\newblock
{\BBOQ}\APACrefatitle {Automatic spectral analysis with missing data}
  {Automatic spectral analysis with missing data}.{\BBCQ}
\newblock
\APACjournalVolNumPages{Digital Signal Processing: A Review
  Journal}{16}{}{754-766}.
\newblock
\begin{APACrefDOI} \doi{10.1016/j.dsp.2006.01.001} \end{APACrefDOI}
\PrintBackRefs{\CurrentBib}

\bibitem [\protect \citeauthoryear {%
M\BPBI L.~Brown%
\ \BBA {} Kros%
}{%
M\BPBI L.~Brown%
\ \BBA {} Kros%
}{%
{\protect \APACyear {2003}}%
}]{%
Brown2003}
\APACinsertmetastar {%
Brown2003}%
\begin{APACrefauthors}%
Brown, M\BPBI L.%
\BCBT {}\ \BBA {} Kros, J\BPBI F.%
\end{APACrefauthors}%
\unskip\
\newblock
\APACrefYearMonthDay{2003}{11}{}.
\newblock
{\BBOQ}\APACrefatitle {Data mining and the impact of missing data} {Data mining
  and the impact of missing data}.{\BBCQ}
\newblock
\APACjournalVolNumPages{Industrial Management \& Data Systems}{103}{}{611-621}.
\newblock
\begin{APACrefURL}
  \url{https://www.emerald.com/insight/content/doi/10.1108/02635570310497657/full/html}
  \end{APACrefURL}
\newblock
\begin{APACrefDOI} \doi{10.1108/02635570310497657} \end{APACrefDOI}
\PrintBackRefs{\CurrentBib}

\bibitem [\protect \citeauthoryear {%
T\BPBI M.~Brown%
\ \BBA {} Christensen-Dalsgaard%
}{%
T\BPBI M.~Brown%
\ \BBA {} Christensen-Dalsgaard%
}{%
{\protect \APACyear {1990}}%
}]{%
Brown1990}
\APACinsertmetastar {%
Brown1990}%
\begin{APACrefauthors}%
Brown, T\BPBI M.%
\BCBT {}\ \BBA {} Christensen-Dalsgaard, J.%
\end{APACrefauthors}%
\unskip\
\newblock
\APACrefYearMonthDay{1990}{}{}.
\newblock
{\BBOQ}\APACrefatitle {A technique for estimating complicated power spectra
  from time series with gaps} {A technique for estimating complicated power
  spectra from time series with gaps}.{\BBCQ}
\newblock
\APACjournalVolNumPages{The Astrophysical Journal}{349}{}{667-674}.
\PrintBackRefs{\CurrentBib}

\bibitem [\protect \citeauthoryear {%
Brownlee%
}{%
Brownlee%
}{%
{\protect \APACyear {2021}}%
}]{%
Brownlee2022}
\APACinsertmetastar {%
Brownlee2022}%
\begin{APACrefauthors}%
Brownlee, J.%
\end{APACrefauthors}%
\unskip\
\newblock
\APACrefYearMonthDay{2021}{}{}.
\newblock
\APACrefbtitle {{How to Choose an Activation Function for Deep Learning}.}
  {{How to Choose an Activation Function for Deep Learning}.}
\newblock
\begin{APACrefURL}
  \url{https://machinelearningmastery.com/choose-an-activation-function-for-deep-learning/}
  \end{APACrefURL}
\newblock
\APACrefnote{{Accessed 23/05/2022}}
\PrintBackRefs{\CurrentBib}

\bibitem [\protect \citeauthoryear {%
Bruno%
\ \BBA {} Carbone%
}{%
Bruno%
\ \BBA {} Carbone%
}{%
{\protect \APACyear {2013}}%
}]{%
BrunoLRSP13}
\APACinsertmetastar {%
BrunoLRSP13}%
\begin{APACrefauthors}%
Bruno, R.%
\BCBT {}\ \BBA {} Carbone, V.%
\end{APACrefauthors}%
\unskip\
\newblock
\APACrefYearMonthDay{2013}{}{}.
\newblock
{\BBOQ}\APACrefatitle {The Solar Wind as a Turbulence Laboratory} {The solar
  wind as a turbulence laboratory}.{\BBCQ}
\newblock
\APACjournalVolNumPages{Living Reviews in Solar Physics}{10}{2}{}.
\newblock
\begin{APACrefURL} \url{http://www.livingreviews.org/lrsp-2013-2}
  \end{APACrefURL}
\newblock
\begin{APACrefDOI} \doi{10.12942/lrsp-2013-2} \end{APACrefDOI}
\PrintBackRefs{\CurrentBib}

\bibitem [\protect \citeauthoryear {%
Burlaga%
}{%
Burlaga%
}{%
{\protect \APACyear {1991}}%
}]{%
burlaga_91}
\APACinsertmetastar {%
burlaga_91}%
\begin{APACrefauthors}%
Burlaga, L.%
\end{APACrefauthors}%
\unskip\
\newblock
\APACrefYearMonthDay{1991}{}{}.
\newblock
{\BBOQ}\APACrefatitle {Intermittent turbulence in the solar wind} {Intermittent
  turbulence in the solar wind}.{\BBCQ}
\newblock
\APACjournalVolNumPages{Journal of Geophysical Research}{96}{A4}{5847-5851}.
\PrintBackRefs{\CurrentBib}

\bibitem [\protect \citeauthoryear {%
Burlaga%
, Florinski%
\BCBL {}\ \BBA {} Ness%
}{%
Burlaga%
\ \protect \BOthers {.}}{%
{\protect \APACyear {2018}}%
}]{%
BurlagaApJ2018}
\APACinsertmetastar {%
BurlagaApJ2018}%
\begin{APACrefauthors}%
Burlaga, L.%
, Florinski, V.%
\BCBL {}\ \BBA {} Ness, N.%
\end{APACrefauthors}%
\unskip\
\newblock
\APACrefYearMonthDay{2018}{}{}.
\newblock
{\BBOQ}\APACrefatitle {Turbulence in the outer heliosheath} {Turbulence in the
  outer heliosheath}.{\BBCQ}
\newblock
\APACjournalVolNumPages{The Astrophysical Journal}{854}{1}{20}.
\PrintBackRefs{\CurrentBib}

\bibitem [\protect \citeauthoryear {%
Caiafa%
, Sun%
, Tanaka%
, Marti-Puig%
\BCBL {}\ \BBA {} Solé-Casals%
}{%
Caiafa%
\ \protect \BOthers {.}}{%
{\protect \APACyear {2021}}%
}]{%
Caiafa2021}
\APACinsertmetastar {%
Caiafa2021}%
\begin{APACrefauthors}%
Caiafa, C\BPBI F.%
, Sun, Z.%
, Tanaka, T.%
, Marti-Puig, P.%
\BCBL {}\ \BBA {} Solé-Casals, J.%
\end{APACrefauthors}%
\unskip\
\newblock
\APACrefYearMonthDay{2021}{}{}.
\newblock
{\BBOQ}\APACrefatitle {Machine Learning Methods with Noisy, Incomplete or Small
  Datasets} {Machine learning methods with noisy, incomplete or small
  datasets}.{\BBCQ}
\newblock
\APACjournalVolNumPages{Applied Sciences}{11}{9}{}.
\newblock
\begin{APACrefURL} \url{https://www.mdpi.com/2076-3417/11/9/4132}
  \end{APACrefURL}
\newblock
\begin{APACrefDOI} \doi{10.3390/app11094132} \end{APACrefDOI}
\PrintBackRefs{\CurrentBib}

\bibitem [\protect \citeauthoryear {%
Camporeale%
}{%
Camporeale%
}{%
{\protect \APACyear {2019}}%
}]{%
camporeale_19}
\APACinsertmetastar {%
camporeale_19}%
\begin{APACrefauthors}%
Camporeale, E.%
\end{APACrefauthors}%
\unskip\
\newblock
\APACrefYearMonthDay{2019}{}{}.
\newblock
{\BBOQ}\APACrefatitle {The Challenge of Machine Learning in Space Weather:
  Nowcasting and Forecasting} {The challenge of machine learning in space
  weather: Nowcasting and forecasting}.{\BBCQ}
\newblock
\APACjournalVolNumPages{Space Weather}{17}{8}{1166-1207}.
\PrintBackRefs{\CurrentBib}

\bibitem [\protect \citeauthoryear {%
Camporeale%
, Carè%
\BCBL {}\ \BBA {} Borovsky%
}{%
Camporeale%
\ \protect \BOthers {.}}{%
{\protect \APACyear {2017}}%
}]{%
Camporeale2017}
\APACinsertmetastar {%
Camporeale2017}%
\begin{APACrefauthors}%
Camporeale, E.%
, Carè, A.%
\BCBL {}\ \BBA {} Borovsky, J\BPBI E.%
\end{APACrefauthors}%
\unskip\
\newblock
\APACrefYearMonthDay{2017}{}{}.
\newblock
{\BBOQ}\APACrefatitle {Classification of Solar Wind With Machine Learning}
  {Classification of solar wind with machine learning}.{\BBCQ}
\newblock
\APACjournalVolNumPages{Journal of Geophysical Research: Space
  Physics}{122}{11}{10,910-10,920}.
\newblock
\begin{APACrefURL}
  \url{https://agupubs.onlinelibrary.wiley.com/doi/abs/10.1002/2017JA024383}
  \end{APACrefURL}
\newblock
\begin{APACrefDOI} \doi{https://doi.org/10.1002/2017JA024383} \end{APACrefDOI}
\PrintBackRefs{\CurrentBib}

\bibitem [\protect \citeauthoryear {%
C.~Chen%
\ \protect \BOthers {.}}{%
C.~Chen%
\ \protect \BOthers {.}}{%
{\protect \APACyear {2020}}%
}]{%
ChenApJS20}
\APACinsertmetastar {%
ChenApJS20}%
\begin{APACrefauthors}%
Chen, C.%
, Bale, S.%
, Bonnell, J.%
, Borovikov, D.%
, Bowen, T.%
, Burgess, D.%
\BDBL {}others%
\end{APACrefauthors}%
\unskip\
\newblock
\APACrefYearMonthDay{2020}{}{}.
\newblock
{\BBOQ}\APACrefatitle {The evolution and role of solar wind turbulence in the
  inner heliosphere} {The evolution and role of solar wind turbulence in the
  inner heliosphere}.{\BBCQ}
\newblock
\APACjournalVolNumPages{The Astrophysical Journal Supplement
  Series}{246}{2}{53}.
\PrintBackRefs{\CurrentBib}

\bibitem [\protect \citeauthoryear {%
Y.~Chen%
, Kopp%
\BCBL {}\ \BBA {} Surry%
}{%
Y.~Chen%
\ \protect \BOthers {.}}{%
{\protect \APACyear {2002}}%
}]{%
chen_2002}
\APACinsertmetastar {%
chen_2002}%
\begin{APACrefauthors}%
Chen, Y.%
, Kopp, G\BPBI A.%
\BCBL {}\ \BBA {} Surry, D.%
\end{APACrefauthors}%
\unskip\
\newblock
\APACrefYearMonthDay{2002}{}{}.
\newblock
{\BBOQ}\APACrefatitle {Interpolation of wind-induced pressure time series with
  an artificial neural network} {Interpolation of wind-induced pressure time
  series with an artificial neural network}.{\BBCQ}
\newblock
\APACjournalVolNumPages{Journal of Wind Engineering and Industrial
  Aerodynamics}{90}{}{589-615}.
\PrintBackRefs{\CurrentBib}

\bibitem [\protect \citeauthoryear {%
Chhiber%
\ \protect \BOthers {.}}{%
Chhiber%
\ \protect \BOthers {.}}{%
{\protect \APACyear {2018}}%
}]{%
Chhiber2018}
\APACinsertmetastar {%
Chhiber2018}%
\begin{APACrefauthors}%
Chhiber, R.%
, Chasapis, A.%
, Bandyopadhyay, R.%
, Parashar, T\BPBI N.%
, Matthaeus, W\BPBI H.%
, Maruca, B.%
\BDBL {}Gershman, D\BPBI J.%
\end{APACrefauthors}%
\unskip\
\newblock
\APACrefYearMonthDay{2018}{12}{}.
\newblock
{\BBOQ}\APACrefatitle {Higher‐order Turbulence Statistics in the Earth's
  Magnetosheath and the Solar Wind using Magnetospheric Multiscale
  Observations} {Higher‐order turbulence statistics in the earth's
  magnetosheath and the solar wind using magnetospheric multiscale
  observations}.{\BBCQ}
\newblock
\APACjournalVolNumPages{Journal of Geophysical Research: Space
  Physics}{123}{}{2018JA025768}.
\newblock
\begin{APACrefURL}
  \url{https://onlinelibrary.wiley.com/doi/abs/10.1029/2018JA025768}
  \end{APACrefURL}
\newblock
\begin{APACrefDOI} \doi{10.1029/2018JA025768} \end{APACrefDOI}
\PrintBackRefs{\CurrentBib}

\bibitem [\protect \citeauthoryear {%
Chollet%
}{%
Chollet%
}{%
{\protect \APACyear {2021}}%
}]{%
Chollet2021}
\APACinsertmetastar {%
Chollet2021}%
\begin{APACrefauthors}%
Chollet, F.%
\end{APACrefauthors}%
\unskip\
\newblock
\APACrefYearMonthDay{2021}{}{}.
\newblock
{\BBOQ}\APACrefatitle {The universal workflow of machine learning} {The
  universal workflow of machine learning}.{\BBCQ}
\newblock
\BIn{} \APACrefbtitle {Deep Learning with {P}ython, Second Edition} {Deep
  learning with {P}ython, second edition}\ (\BCHAP~6).
\newblock
\APACaddressPublisher{}{Manning Publications}.
\PrintBackRefs{\CurrentBib}

\bibitem [\protect \citeauthoryear {%
{Coleman Jr, P~J}%
}{%
{Coleman Jr, P~J}%
}{%
{\protect \APACyear {1968}}%
}]{%
ColemanApJ68}
\APACinsertmetastar {%
ColemanApJ68}%
\begin{APACrefauthors}%
{Coleman Jr, P~J}.%
\end{APACrefauthors}%
\unskip\
\newblock
\APACrefYearMonthDay{1968}{{\APACmonth{08}}}{}.
\newblock
{\BBOQ}\APACrefatitle {{Turbulence, Viscosity, and Dissipation in the
  Solar-Wind Plasma}} {{Turbulence, Viscosity, and Dissipation in the
  Solar-Wind Plasma}}.{\BBCQ}
\newblock
\APACjournalVolNumPages{The Astrophysical Journal}{153}{}{371-+}.
\newblock
\begin{APACrefDOI} \doi{10.1086/149674} \end{APACrefDOI}
\PrintBackRefs{\CurrentBib}

\bibitem [\protect \citeauthoryear {%
Comerford%
, Kougioumtzoglou%
\BCBL {}\ \BBA {} Beer%
}{%
Comerford%
\ \protect \BOthers {.}}{%
{\protect \APACyear {2015}}%
}]{%
Comerford2015}
\APACinsertmetastar {%
Comerford2015}%
\begin{APACrefauthors}%
Comerford, L.%
, Kougioumtzoglou, I\BPBI A.%
\BCBL {}\ \BBA {} Beer, M.%
\end{APACrefauthors}%
\unskip\
\newblock
\APACrefYearMonthDay{2015}{1}{}.
\newblock
{\BBOQ}\APACrefatitle {An artificial neural network approach for stochastic
  process power spectrum estimation subject to missing data} {An artificial
  neural network approach for stochastic process power spectrum estimation
  subject to missing data}.{\BBCQ}
\newblock
\APACjournalVolNumPages{Structural Safety}{52}{}{150-160}.
\newblock
\begin{APACrefDOI} \doi{10.1016/j.strusafe.2014.10.001} \end{APACrefDOI}
\PrintBackRefs{\CurrentBib}

\bibitem [\protect \citeauthoryear {%
de Souza~Echer%
\ \protect \BOthers {.}}{%
de Souza~Echer%
\ \protect \BOthers {.}}{%
{\protect \APACyear {2021}}%
}]{%
deSouzaEcher2021}
\APACinsertmetastar {%
deSouzaEcher2021}%
\begin{APACrefauthors}%
de Souza~Echer, M\BPBI P.%
, Echer, E.%
, Domingues, M\BPBI O.%
, Mendes, O.%
, Seo, R\BPBI T.%
\BCBL {}\ \BBA {} Gonzalez, W.%
\end{APACrefauthors}%
\unskip\
\newblock
\APACrefYearMonthDay{2021}{}{}.
\newblock
{\BBOQ}\APACrefatitle {Wavelet analysis of low frequency magnetic field
  fluctuations in the {Jupiter's} magnetotail} {Wavelet analysis of low
  frequency magnetic field fluctuations in the {Jupiter's} magnetotail}.{\BBCQ}
\newblock
\APACjournalVolNumPages{Advances in Space Research}{68}{}{246-258}.
\newblock
\begin{APACrefDOI} \doi{10.1016/j.asr.2021.03.003} \end{APACrefDOI}
\PrintBackRefs{\CurrentBib}

\bibitem [\protect \citeauthoryear {%
Emmanuel%
\ \protect \BOthers {.}}{%
Emmanuel%
\ \protect \BOthers {.}}{%
{\protect \APACyear {2021}}%
}]{%
Emmanuel2021}
\APACinsertmetastar {%
Emmanuel2021}%
\begin{APACrefauthors}%
Emmanuel, T.%
, Maupong, T.%
, Mpoeleng, D.%
, Semong, T.%
, Mphago, B.%
\BCBL {}\ \BBA {} Tabona, O.%
\end{APACrefauthors}%
\unskip\
\newblock
\APACrefYearMonthDay{2021}{}{}.
\newblock
{\BBOQ}\APACrefatitle {A survey on missing data in machine learning} {A survey
  on missing data in machine learning}.{\BBCQ}
\newblock
\APACjournalVolNumPages{Journal of Big Data}{8}{}{140}.
\newblock
\begin{APACrefURL}
  \url{https://journalofbigdata.springeropen.com/articles/10.1186/s40537-021-00516-9}
  \end{APACrefURL}
\newblock
\begin{APACrefDOI} \doi{10.1186/s40537-021-00516-9} \end{APACrefDOI}
\PrintBackRefs{\CurrentBib}

\bibitem [\protect \citeauthoryear {%
Finch%
\ \BBA {} Lockwood%
}{%
Finch%
\ \BBA {} Lockwood%
}{%
{\protect \APACyear {2007}}%
}]{%
Finch2007}
\APACinsertmetastar {%
Finch2007}%
\begin{APACrefauthors}%
Finch, I.%
\BCBT {}\ \BBA {} Lockwood, M.%
\end{APACrefauthors}%
\unskip\
\newblock
\APACrefYearMonthDay{2007}{}{}.
\newblock
{\BBOQ}\APACrefatitle {Solar wind-magnetosphere coupling functions on
  timescales of 1 day to 1 year} {Solar wind-magnetosphere coupling functions
  on timescales of 1 day to 1 year}.{\BBCQ}
\newblock
\APACjournalVolNumPages{Annales Geophysicae}{25}{}{495-506}.
\newblock
\begin{APACrefURL} \url{www.ann-geophys.net/25/495/2007/} \end{APACrefURL}
\newblock
\APACrefnote{See also Lockwood's 2019 article}
\PrintBackRefs{\CurrentBib}

\bibitem [\protect \citeauthoryear {%
Fox%
\ \protect \BOthers {.}}{%
Fox%
\ \protect \BOthers {.}}{%
{\protect \APACyear {2016}}%
}]{%
Fox2016}
\APACinsertmetastar {%
Fox2016}%
\begin{APACrefauthors}%
Fox, N\BPBI J.%
, Velli, M\BPBI C.%
, Bale, S\BPBI D.%
, Decker, R.%
, Driesman, A.%
, Howard, R\BPBI A.%
\BDBL {}Szabo, A.%
\end{APACrefauthors}%
\unskip\
\newblock
\APACrefYearMonthDay{2016}{12}{}.
\newblock
{\BBOQ}\APACrefatitle {The Solar Probe Plus Mission: Humanity’s First Visit
  to Our Star} {The solar probe plus mission: Humanity’s first visit to our
  star}.{\BBCQ}
\newblock
\APACjournalVolNumPages{Space Science Reviews}{204}{}{7-48}.
\newblock
\begin{APACrefURL} \url{http://link.springer.com/10.1007/s11214-015-0211-6}
  \end{APACrefURL}
\newblock
\begin{APACrefDOI} \doi{10.1007/s11214-015-0211-6} \end{APACrefDOI}
\PrintBackRefs{\CurrentBib}

\bibitem [\protect \citeauthoryear {%
Fraternale%
, Pogorelov%
, Richardson%
\BCBL {}\ \BBA {} Tordella%
}{%
Fraternale%
\ \protect \BOthers {.}}{%
{\protect \APACyear {2019}}%
}]{%
Fraternale2019}
\APACinsertmetastar {%
Fraternale2019}%
\begin{APACrefauthors}%
Fraternale, F.%
, Pogorelov, N\BPBI V.%
, Richardson, J\BPBI D.%
\BCBL {}\ \BBA {} Tordella, D.%
\end{APACrefauthors}%
\unskip\
\newblock
\APACrefYearMonthDay{2019}{2}{}.
\newblock
{\BBOQ}\APACrefatitle {Magnetic Turbulence Spectra and Intermittency in the
  Heliosheath and in the Local Interstellar Medium} {Magnetic turbulence
  spectra and intermittency in the heliosheath and in the local interstellar
  medium}.{\BBCQ}
\newblock
\APACjournalVolNumPages{The Astrophysical Journal}{872}{}{40}.
\newblock
\begin{APACrefDOI} \doi{10.3847/1538-4357/aafd30} \end{APACrefDOI}
\PrintBackRefs{\CurrentBib}

\bibitem [\protect \citeauthoryear {%
Frick%
, Grossmann%
\BCBL {}\ \BBA {} Tchamitchian%
}{%
Frick%
\ \protect \BOthers {.}}{%
{\protect \APACyear {1998}}%
}]{%
Frick1998}
\APACinsertmetastar {%
Frick1998}%
\begin{APACrefauthors}%
Frick, P.%
, Grossmann, A.%
\BCBL {}\ \BBA {} Tchamitchian, P.%
\end{APACrefauthors}%
\unskip\
\newblock
\APACrefYearMonthDay{1998}{}{}.
\newblock
{\BBOQ}\APACrefatitle {Wavelet analysis of signals with gaps} {Wavelet analysis
  of signals with gaps}.{\BBCQ}
\newblock
\APACjournalVolNumPages{Journal of Mathematical Physics}{39}{}{4091-4107}.
\newblock
\begin{APACrefDOI} \doi{10.1063/1.532485} \end{APACrefDOI}
\PrintBackRefs{\CurrentBib}

\bibitem [\protect \citeauthoryear {%
Friedrich%
, Gallon%
, Pumir%
\BCBL {}\ \BBA {} Grauer%
}{%
Friedrich%
\ \protect \BOthers {.}}{%
{\protect \APACyear {2020}}%
}]{%
Friedrich2020}
\APACinsertmetastar {%
Friedrich2020}%
\begin{APACrefauthors}%
Friedrich, J.%
, Gallon, S.%
, Pumir, A.%
\BCBL {}\ \BBA {} Grauer, R.%
\end{APACrefauthors}%
\unskip\
\newblock
\APACrefYearMonthDay{2020}{}{}.
\newblock
{\BBOQ}\APACrefatitle {Stochastic Interpolation of Sparsely Sampled Time Series
  via Multipoint Fractional Brownian Bridges} {Stochastic interpolation of
  sparsely sampled time series via multipoint fractional brownian
  bridges}.{\BBCQ}
\newblock
\APACjournalVolNumPages{Physical Review Letters}{125}{}{}.
\newblock
\begin{APACrefDOI} \doi{10.1103/PhysRevLett.125.170602} \end{APACrefDOI}
\PrintBackRefs{\CurrentBib}

\bibitem [\protect \citeauthoryear {%
Frisch%
}{%
Frisch%
}{%
{\protect \APACyear {1995}}%
}]{%
Frisch1995}
\APACinsertmetastar {%
Frisch1995}%
\begin{APACrefauthors}%
Frisch, U.%
\end{APACrefauthors}%
\unskip\
\newblock
\APACrefYear{1995}.
\newblock
\APACrefbtitle {Turbulence: The Legacy of A. N. Kolmogorov} {Turbulence: The
  legacy of a. n. kolmogorov}.
\newblock
\APACaddressPublisher{}{Cambridge University Press}.
\newblock
\begin{APACrefURL} \url{https://books.google.co.nz/books?id=-JcjT4wYgfgC}
  \end{APACrefURL}
\PrintBackRefs{\CurrentBib}

\bibitem [\protect \citeauthoryear {%
Gallana%
\ \protect \BOthers {.}}{%
Gallana%
\ \protect \BOthers {.}}{%
{\protect \APACyear {2016}}%
}]{%
Gallana2016}
\APACinsertmetastar {%
Gallana2016}%
\begin{APACrefauthors}%
Gallana, L.%
, Fraternale, F.%
, Iovieno, M.%
, Fosson, S\BPBI M.%
, Magli, E.%
, Opher, M.%
\BDBL {}Tordella, D.%
\end{APACrefauthors}%
\unskip\
\newblock
\APACrefYearMonthDay{2016}{}{}.
\newblock
{\BBOQ}\APACrefatitle {Voyager 2 solar plasma and magnetic field spectral
  analysis for intermediate data sparsity} {Voyager 2 solar plasma and magnetic
  field spectral analysis for intermediate data sparsity}.{\BBCQ}
\newblock
\APACjournalVolNumPages{Journal of Geophysical Research: Space
  Physics}{121}{}{3905--3919}.
\PrintBackRefs{\CurrentBib}

\bibitem [\protect \citeauthoryear {%
Harvey%
\ \BBA {} Pierse%
}{%
Harvey%
\ \BBA {} Pierse%
}{%
{\protect \APACyear {1984}}%
}]{%
Harvey1984}
\APACinsertmetastar {%
Harvey1984}%
\begin{APACrefauthors}%
Harvey, A\BPBI C.%
\BCBT {}\ \BBA {} Pierse, R\BPBI G.%
\end{APACrefauthors}%
\unskip\
\newblock
\APACrefYearMonthDay{1984}{}{}.
\newblock
{\BBOQ}\APACrefatitle {Estimating Missing Observations in Economic Time Series}
  {Estimating missing observations in economic time series}.{\BBCQ}
\newblock
\APACjournalVolNumPages{Source: Journal of the American Statistical
  Association}{79}{}{125-131}.
\PrintBackRefs{\CurrentBib}

\bibitem [\protect \citeauthoryear {%
He%
, Tu%
, Marsch%
\BCBL {}\ \BBA {} Yao%
}{%
He%
\ \protect \BOthers {.}}{%
{\protect \APACyear {2011}}%
}]{%
He_2011}
\APACinsertmetastar {%
He_2011}%
\begin{APACrefauthors}%
He, J.%
, Tu, C.%
, Marsch, E.%
\BCBL {}\ \BBA {} Yao, S.%
\end{APACrefauthors}%
\unskip\
\newblock
\APACrefYearMonthDay{2011}{dec}{}.
\newblock
{\BBOQ}\APACrefatitle {{DO} {OBLIQUE} {ALFV}{\'{E}}N/{ION}-{CYCLOTRON} {OR}
  {FAST}-{MODE}/{WHISTLER} {WAVES} {DOMINATE} {THE} {DISSIPATION} {OF} {SOLAR}
  {WIND} {TURBULENCE} {NEAR} {THE} {PROTON} {INERTIAL} {LENGTH}?} {{DO}
  {OBLIQUE} {ALFV}{\'{e}}n/{ION}-{CYCLOTRON} {OR} {FAST}-{MODE}/{WHISTLER}
  {WAVES} {DOMINATE} {THE} {DISSIPATION} {OF} {SOLAR} {WIND} {TURBULENCE}
  {NEAR} {THE} {PROTON} {INERTIAL} {LENGTH}?}{\BBCQ}
\newblock
\APACjournalVolNumPages{The Astrophysical Journal}{745}{1}{L8}.
\newblock
\begin{APACrefURL} \url{https://doi.org/10.1088/2041-8205/745/1/l8}
  \end{APACrefURL}
\newblock
\begin{APACrefDOI} \doi{10.1088/2041-8205/745/1/l8} \end{APACrefDOI}
\PrintBackRefs{\CurrentBib}

\bibitem [\protect \citeauthoryear {%
Hu%
\ \protect \BOthers {.}}{%
Hu%
\ \protect \BOthers {.}}{%
{\protect \APACyear {2020}}%
}]{%
Hu_2020}
\APACinsertmetastar {%
Hu_2020}%
\begin{APACrefauthors}%
Hu, A.%
, Sisti, M.%
, Finelli, F.%
, Califano, F.%
, Dargent, J.%
, Faganello, M.%
\BDBL {}Teunissen, J.%
\end{APACrefauthors}%
\unskip\
\newblock
\APACrefYearMonthDay{2020}{}{}.
\newblock
{\BBOQ}\APACrefatitle {Identifying Magnetic Reconnection in 2D Hybrid Vlasov
  Maxwell Simulations with Convolutional Neural Networks} {Identifying magnetic
  reconnection in 2d hybrid vlasov maxwell simulations with convolutional
  neural networks}.{\BBCQ}
\newblock
\APACjournalVolNumPages{The Astrophysical Journal}{900}{1}{86}.
\PrintBackRefs{\CurrentBib}

\bibitem [\protect \citeauthoryear {%
Isaacs%
, Tessein%
\BCBL {}\ \BBA {} Matthaeus%
}{%
Isaacs%
\ \protect \BOthers {.}}{%
{\protect \APACyear {2015}}%
}]{%
IsaacsJGR15}
\APACinsertmetastar {%
IsaacsJGR15}%
\begin{APACrefauthors}%
Isaacs, J.%
, Tessein, J.%
\BCBL {}\ \BBA {} Matthaeus, W.%
\end{APACrefauthors}%
\unskip\
\newblock
\APACrefYearMonthDay{2015}{}{}.
\newblock
{\BBOQ}\APACrefatitle {Systematic averaging interval effects on solar wind
  statistics} {Systematic averaging interval effects on solar wind
  statistics}.{\BBCQ}
\newblock
\APACjournalVolNumPages{Journal of Geophysical Research: Space
  Physics}{120}{2}{868--879}.
\PrintBackRefs{\CurrentBib}

\bibitem [\protect \citeauthoryear {%
Jagarlamudi%
, de Wit%
, Krasnoselskikh%
\BCBL {}\ \BBA {} Maksimovic%
}{%
Jagarlamudi%
\ \protect \BOthers {.}}{%
{\protect \APACyear {2019}}%
}]{%
JagarlamudiApJ19}
\APACinsertmetastar {%
JagarlamudiApJ19}%
\begin{APACrefauthors}%
Jagarlamudi, V\BPBI K.%
, de Wit, T\BPBI D.%
, Krasnoselskikh, V.%
\BCBL {}\ \BBA {} Maksimovic, M.%
\end{APACrefauthors}%
\unskip\
\newblock
\APACrefYearMonthDay{2019}{}{}.
\newblock
{\BBOQ}\APACrefatitle {Inherentness of Non-stationarity in Solar Wind}
  {Inherentness of non-stationarity in solar wind}.{\BBCQ}
\newblock
\APACjournalVolNumPages{The Astrophysical Journal}{871}{1}{68}.
\PrintBackRefs{\CurrentBib}

\bibitem [\protect \citeauthoryear {%
Jang%
\ \protect \BOthers {.}}{%
Jang%
\ \protect \BOthers {.}}{%
{\protect \APACyear {2020}}%
}]{%
jang_2020}
\APACinsertmetastar {%
jang_2020}%
\begin{APACrefauthors}%
Jang, J.%
, Choi, K.%
, Roh, H.%
, Son, S.%
, Hong, C.%
, Kim, E.%
\BDBL {}Yoon, D.%
\end{APACrefauthors}%
\unskip\
\newblock
\APACrefYearMonthDay{2020}{}{}.
\newblock
{\BBOQ}\APACrefatitle {Deep Learning Approach for Imputation of Missing Values
  in Actigraphy Data: Algorithm Development Study} {Deep learning approach for
  imputation of missing values in actigraphy data: Algorithm development
  study}.{\BBCQ}
\newblock
\APACjournalVolNumPages{JMIR Mhealth Uhealth}{8}{}{}.
\PrintBackRefs{\CurrentBib}

\bibitem [\protect \citeauthoryear {%
Jiao%
\ \protect \BOthers {.}}{%
Jiao%
\ \protect \BOthers {.}}{%
{\protect \APACyear {2020}}%
}]{%
Jiao2020}
\APACinsertmetastar {%
Jiao2020}%
\begin{APACrefauthors}%
Jiao, Z.%
, Sun, H.%
, Wang, X.%
, Manchester, W.%
, Gombosi, T.%
, Hero, A.%
\BCBL {}\ \BBA {} Chen, Y.%
\end{APACrefauthors}%
\unskip\
\newblock
\APACrefYearMonthDay{2020}{}{}.
\newblock
{\BBOQ}\APACrefatitle {Solar Flare Intensity Prediction With Machine Learning
  Models} {Solar flare intensity prediction with machine learning
  models}.{\BBCQ}
\newblock
\APACjournalVolNumPages{Space Weather}{18}{7}{e2020SW002440}.
\newblock
\begin{APACrefURL}
  \url{https://agupubs.onlinelibrary.wiley.com/doi/abs/10.1029/2020SW002440}
  \end{APACrefURL}
\newblock
\APACrefnote{e2020SW002440 10.1029/2020SW002440}
\newblock
\begin{APACrefDOI} \doi{https://doi.org/10.1029/2020SW002440} \end{APACrefDOI}
\PrintBackRefs{\CurrentBib}

\bibitem [\protect \citeauthoryear {%
Kasper%
\ \protect \BOthers {.}}{%
Kasper%
\ \protect \BOthers {.}}{%
{\protect \APACyear {2019}}%
}]{%
KasperNature19}
\APACinsertmetastar {%
KasperNature19}%
\begin{APACrefauthors}%
Kasper, J\BPBI C.%
, Bale, S\BPBI D.%
, Belcher, J\BPBI W.%
, Berthomier, M.%
, Case, A\BPBI W.%
, Chandran, B\BPBI D.%
\BDBL {}others%
\end{APACrefauthors}%
\unskip\
\newblock
\APACrefYearMonthDay{2019}{}{}.
\newblock
{\BBOQ}\APACrefatitle {Alfv{\'e}nic velocity spikes and rotational flows in the
  near-Sun solar wind} {Alfv{\'e}nic velocity spikes and rotational flows in
  the near-sun solar wind}.{\BBCQ}
\newblock
\APACjournalVolNumPages{Nature}{576}{7786}{228--231}.
\PrintBackRefs{\CurrentBib}

\bibitem [\protect \citeauthoryear {%
Kim%
\ \protect \BOthers {.}}{%
Kim%
\ \protect \BOthers {.}}{%
{\protect \APACyear {2019}}%
}]{%
Kim2019}
\APACinsertmetastar {%
Kim2019}%
\begin{APACrefauthors}%
Kim, T.%
, Park, E.%
, Lee, H.%
, Moon, Y\BHBI J.%
, Bae, S\BHBI H.%
, Lim, D.%
\BDBL {}Cho, K\BHBI S.%
\end{APACrefauthors}%
\unskip\
\newblock
\APACrefYearMonthDay{2019}{}{}.
\newblock
{\BBOQ}\APACrefatitle {Solar farside magnetograms from deep learning analysis
  of STEREO/EUVI data} {Solar farside magnetograms from deep learning analysis
  of stereo/euvi data}.{\BBCQ}
\newblock
\APACjournalVolNumPages{Nature Astronomy}{3}{5}{397-400}.
\newblock
\begin{APACrefDOI} \doi{10.1038/s41550-019-0711-5} \end{APACrefDOI}
\PrintBackRefs{\CurrentBib}

\bibitem [\protect \citeauthoryear {%
Klein%
, Howes%
, TenBarge%
\BCBL {}\ \BBA {} Podesta%
}{%
Klein%
\ \protect \BOthers {.}}{%
{\protect \APACyear {2014}}%
}]{%
Klein_2014}
\APACinsertmetastar {%
Klein_2014}%
\begin{APACrefauthors}%
Klein, K\BPBI G.%
, Howes, G\BPBI G.%
, TenBarge, J\BPBI M.%
\BCBL {}\ \BBA {} Podesta, J\BPBI J.%
\end{APACrefauthors}%
\unskip\
\newblock
\APACrefYearMonthDay{2014}{apr}{}.
\newblock
{\BBOQ}\APACrefatitle {{PHYSICAL} {INTERPRETATION} {OF} {THE}
  {ANGLE}-{DEPENDENT} {MAGNETIC} {HELICITY} {SPECTRUM} {IN} {THE} {SOLAR}
  {WIND}: {THE} {NATURE} {OF} {TURBULENT} {FLUCTUATIONS} {NEAR} {THE} {PROTON}
  {GYRORADIUS} {SCALE}} {{PHYSICAL} {INTERPRETATION} {OF} {THE}
  {ANGLE}-{DEPENDENT} {MAGNETIC} {HELICITY} {SPECTRUM} {IN} {THE} {SOLAR}
  {WIND}: {THE} {NATURE} {OF} {TURBULENT} {FLUCTUATIONS} {NEAR} {THE} {PROTON}
  {GYRORADIUS} {SCALE}}.{\BBCQ}
\newblock
\APACjournalVolNumPages{The Astrophysical Journal}{785}{2}{138}.
\newblock
\begin{APACrefURL} \url{https://doi.org/10.1088/0004-637x/785/2/138}
  \end{APACrefURL}
\newblock
\begin{APACrefDOI} \doi{10.1088/0004-637x/785/2/138} \end{APACrefDOI}
\PrintBackRefs{\CurrentBib}

\bibitem [\protect \citeauthoryear {%
Kobayashi%
, Ozturk%
, Connor%
\BCBL {}\ \BBA {} Keesee%
}{%
Kobayashi%
\ \protect \BOthers {.}}{%
{\protect \APACyear {2021}}%
}]{%
Kobayashi2021}
\APACinsertmetastar {%
Kobayashi2021}%
\begin{APACrefauthors}%
Kobayashi, J.%
, Ozturk, D\BPBI S.%
, Connor, H\BPBI K.%
\BCBL {}\ \BBA {} Keesee, A\BPBI M.%
\end{APACrefauthors}%
\unskip\
\newblock
\APACrefYearMonthDay{2021}{}{}.
\newblock
{\BBOQ}\APACrefatitle {Machine Learning Models as an Alternative to Standard
  Interpolation Techniques for Estimating {OMNI} Data Gaps} {Machine learning
  models as an alternative to standard interpolation techniques for estimating
  {OMNI} data gaps}.{\BBCQ}
\newblock
\BIn{} \APACrefbtitle {{AGU} Fall Meeting 2021.} {{AGU} fall meeting 2021.}
\PrintBackRefs{\CurrentBib}

\bibitem [\protect \citeauthoryear {%
Kondrashov%
, Shprits%
\BCBL {}\ \BBA {} Ghil%
}{%
Kondrashov%
\ \protect \BOthers {.}}{%
{\protect \APACyear {2010}}%
}]{%
Kondrashov2010}
\APACinsertmetastar {%
Kondrashov2010}%
\begin{APACrefauthors}%
Kondrashov, D.%
, Shprits, Y.%
\BCBL {}\ \BBA {} Ghil, M.%
\end{APACrefauthors}%
\unskip\
\newblock
\APACrefYearMonthDay{2010}{8}{}.
\newblock
{\BBOQ}\APACrefatitle {Gap filling of solar wind data by singular spectrum
  analysis} {Gap filling of solar wind data by singular spectrum
  analysis}.{\BBCQ}
\newblock
\APACjournalVolNumPages{Geophysical Research Letters}{37}{}{}.
\newblock
\begin{APACrefDOI} \doi{10.1029/2010GL044138} \end{APACrefDOI}
\PrintBackRefs{\CurrentBib}

\bibitem [\protect \citeauthoryear {%
Lockwood%
\ \protect \BOthers {.}}{%
Lockwood%
\ \protect \BOthers {.}}{%
{\protect \APACyear {2019}}%
}]{%
Lockwood2019}
\APACinsertmetastar {%
Lockwood2019}%
\begin{APACrefauthors}%
Lockwood, M.%
, Bentley, S\BPBI N.%
, Owens, M\BPBI J.%
, Barnard, L\BPBI A.%
, Scott, C\BPBI J.%
, Watt, C\BPBI E.%
\BCBL {}\ \BBA {} Allanson, O.%
\end{APACrefauthors}%
\unskip\
\newblock
\APACrefYearMonthDay{2019}{1}{}.
\newblock
{\BBOQ}\APACrefatitle {The Development of a Space Climatology: 1. Solar Wind
  Magnetosphere Coupling as a Function of Timescale and the Effect of Data
  Gaps} {The development of a space climatology: 1. solar wind magnetosphere
  coupling as a function of timescale and the effect of data gaps}.{\BBCQ}
\newblock
\APACjournalVolNumPages{Space Weather}{17}{}{133-156}.
\newblock
\begin{APACrefDOI} \doi{10.1029/2018SW001856} \end{APACrefDOI}
\PrintBackRefs{\CurrentBib}

\bibitem [\protect \citeauthoryear {%
Ludwig%
\ \BBA {} Taylor%
}{%
Ludwig%
\ \BBA {} Taylor%
}{%
{\protect \APACyear {2016}}%
}]{%
Ludwig2016}
\APACinsertmetastar {%
Ludwig2016}%
\begin{APACrefauthors}%
Ludwig, R.%
\BCBT {}\ \BBA {} Taylor, J.%
\end{APACrefauthors}%
\unskip\
\newblock
\APACrefYearMonthDay{2016}{}{}.
\newblock
{\BBOQ}\APACrefatitle {Voyager Telecommunications} {Voyager
  telecommunications}.{\BBCQ}
\newblock
\BIn{} \APACrefbtitle {Deep Space Communications} {Deep space communications}\
  (\BPG~37-77).
\newblock
\APACaddressPublisher{}{John Wiley \& Sons, Ltd}.
\newblock
\begin{APACrefDOI} \doi{https://doi.org/10.1002/9781119169079.ch3}
  \end{APACrefDOI}
\PrintBackRefs{\CurrentBib}

\bibitem [\protect \citeauthoryear {%
Luo%
, Cai%
, Zhang%
, Xu%
\BCBL {}\ \BBA {} Xiaojie%
}{%
Luo%
\ \protect \BOthers {.}}{%
{\protect \APACyear {2018}}%
}]{%
Luo2018}
\APACinsertmetastar {%
Luo2018}%
\begin{APACrefauthors}%
Luo, Y.%
, Cai, X.%
, Zhang, Y.%
, Xu, J.%
\BCBL {}\ \BBA {} Xiaojie, Y.%
\end{APACrefauthors}%
\unskip\
\newblock
\APACrefYearMonthDay{2018}{}{}.
\newblock
{\BBOQ}\APACrefatitle {Multivariate Time Series Imputation with Generative
  Adversarial Networks} {Multivariate time series imputation with generative
  adversarial networks}.{\BBCQ}
\newblock
\BIn{} S.~Bengio, H.~Wallach, H.~Larochelle, K.~Grauman, N.~Cesa-Bianchi\BCBL
  {}\ \BBA {} R.~Garnett\ (\BEDS), \APACrefbtitle {Advances in Neural
  Information Processing Systems} {Advances in neural information processing
  systems}\ (\BVOL~31).
\newblock
\APACaddressPublisher{}{Curran Associates, Inc.}
\PrintBackRefs{\CurrentBib}

\bibitem [\protect \citeauthoryear {%
Magrini%
, Domingues%
\BCBL {}\ \BBA {} Mendes%
}{%
Magrini%
\ \protect \BOthers {.}}{%
{\protect \APACyear {2017}}%
}]{%
magrini_2017}
\APACinsertmetastar {%
magrini_2017}%
\begin{APACrefauthors}%
Magrini, L\BPBI A.%
, Domingues, M\BPBI O.%
\BCBL {}\ \BBA {} Mendes, O.%
\end{APACrefauthors}%
\unskip\
\newblock
\APACrefYearMonthDay{2017}{}{}.
\newblock
{\BBOQ}\APACrefatitle {On the Effects of Gaps and Uses of Approximation
  Functions on the Time-Scale Signal Analysis: A Case Study Based on Space
  Geophysical Events} {On the effects of gaps and uses of approximation
  functions on the time-scale signal analysis: A case study based on space
  geophysical events}.{\BBCQ}
\newblock
\APACjournalVolNumPages{Brazilian Journal of Physics}{47}{}{167-181}.
\PrintBackRefs{\CurrentBib}

\bibitem [\protect \citeauthoryear {%
Makarynskyy%
, Makarynska%
, E%
\BCBL {}\ \BBA {} A%
}{%
Makarynskyy%
\ \protect \BOthers {.}}{%
{\protect \APACyear {2005}}%
}]{%
Makarynskyy2005}
\APACinsertmetastar {%
Makarynskyy2005}%
\begin{APACrefauthors}%
Makarynskyy, O.%
, Makarynska, D.%
, E, R.%
\BCBL {}\ \BBA {} A, G.%
\end{APACrefauthors}%
\unskip\
\newblock
\APACrefYearMonthDay{2005}{}{}.
\newblock
{\BBOQ}\APACrefatitle {Filling gaps in wave records with artificial neural
  networks} {Filling gaps in wave records with artificial neural
  networks}.{\BBCQ}
\newblock
\APACjournalVolNumPages{International Maritime Association of the
  Mediterranean. International Congress (12th : 2005 : Lisbon,
  Portugal)}{}{}{}.
\PrintBackRefs{\CurrentBib}

\bibitem [\protect \citeauthoryear {%
Marsch%
}{%
Marsch%
}{%
{\protect \APACyear {2006}}%
}]{%
MarschLRSP06}
\APACinsertmetastar {%
MarschLRSP06}%
\begin{APACrefauthors}%
Marsch, E.%
\end{APACrefauthors}%
\unskip\
\newblock
\APACrefYearMonthDay{2006}{}{}.
\newblock
{\BBOQ}\APACrefatitle {Kinetic Physics of the Solar Corona and Solar Wind}
  {Kinetic physics of the solar corona and solar wind}.{\BBCQ}
\newblock
\APACjournalVolNumPages{Living Reviews in Solar Physics}{3}{1}{}.
\newblock
\begin{APACrefURL} \url{http://www.livingreviews.org/lrsp-2006-1}
  \end{APACrefURL}
\PrintBackRefs{\CurrentBib}

\bibitem [\protect \citeauthoryear {%
Matthaeus%
\ \BBA {} Goldstein%
}{%
Matthaeus%
\ \BBA {} Goldstein%
}{%
{\protect \APACyear {1982}}%
}]{%
MatthaeusJGR82}
\APACinsertmetastar {%
MatthaeusJGR82}%
\begin{APACrefauthors}%
Matthaeus, W.%
\BCBT {}\ \BBA {} Goldstein, M.%
\end{APACrefauthors}%
\unskip\
\newblock
\APACrefYearMonthDay{1982}{}{}.
\newblock
{\BBOQ}\APACrefatitle {{Measurement of the rugged invariants of
  magnetohydrodynamic turbulence in the solar wind}} {{Measurement of the
  rugged invariants of magnetohydrodynamic turbulence in the solar
  wind}}.{\BBCQ}
\newblock
\APACjournalVolNumPages{Journal of Geophysical Research}{87}{A8}{6011--6028}.
\PrintBackRefs{\CurrentBib}

\bibitem [\protect \citeauthoryear {%
Munteanu%
, Negrea%
, Echim%
\BCBL {}\ \BBA {} Mursula%
}{%
Munteanu%
\ \protect \BOthers {.}}{%
{\protect \APACyear {2016}}%
}]{%
Munteanu2016}
\APACinsertmetastar {%
Munteanu2016}%
\begin{APACrefauthors}%
Munteanu, C.%
, Negrea, C.%
, Echim, M.%
\BCBL {}\ \BBA {} Mursula, K.%
\end{APACrefauthors}%
\unskip\
\newblock
\APACrefYearMonthDay{2016}{4}{}.
\newblock
{\BBOQ}\APACrefatitle {Effect of data gaps: Comparison of different spectral
  analysis methods} {Effect of data gaps: Comparison of different spectral
  analysis methods}.{\BBCQ}
\newblock
\APACjournalVolNumPages{Annales Geophysicae}{34}{}{437-449}.
\newblock
\begin{APACrefDOI} \doi{10.5194/angeo-34-437-2016} \end{APACrefDOI}
\PrintBackRefs{\CurrentBib}

\bibitem [\protect \citeauthoryear {%
Opher%
\ \protect \BOthers {.}}{%
Opher%
\ \protect \BOthers {.}}{%
{\protect \APACyear {2011}}%
}]{%
OpherApJ11}
\APACinsertmetastar {%
OpherApJ11}%
\begin{APACrefauthors}%
Opher, M.%
, Drake, J.%
, Swisdak, M.%
, Schoeffler, K.%
, Richardson, J.%
, Decker, R.%
\BCBL {}\ \BBA {} Toth, G.%
\end{APACrefauthors}%
\unskip\
\newblock
\APACrefYearMonthDay{2011}{}{}.
\newblock
{\BBOQ}\APACrefatitle {Is the magnetic field in the heliosheath laminar or a
  turbulent sea of bubbles?} {Is the magnetic field in the heliosheath laminar
  or a turbulent sea of bubbles?}{\BBCQ}
\newblock
\APACjournalVolNumPages{The Astrophysical Journal}{734}{1}{71}.
\PrintBackRefs{\CurrentBib}

\bibitem [\protect \citeauthoryear {%
Oran%
\ \protect \BOthers {.}}{%
Oran%
\ \protect \BOthers {.}}{%
{\protect \APACyear {2013}}%
}]{%
oran2013}
\APACinsertmetastar {%
oran2013}%
\begin{APACrefauthors}%
Oran, R.%
, Van~der Holst, B.%
, Landi, E.%
, Jin, M.%
, Sokolov, I.%
\BCBL {}\ \BBA {} Gombosi, T.%
\end{APACrefauthors}%
\unskip\
\newblock
\APACrefYearMonthDay{2013}{}{}.
\newblock
{\BBOQ}\APACrefatitle {A global wave-driven magnetohydrodynamic solar model
  with a unified treatment of open and closed magnetic field topologies} {A
  global wave-driven magnetohydrodynamic solar model with a unified treatment
  of open and closed magnetic field topologies}.{\BBCQ}
\newblock
\APACjournalVolNumPages{The Astrophysical Journal}{778}{2}{176}.
\PrintBackRefs{\CurrentBib}

\bibitem [\protect \citeauthoryear {%
Parashar%
\ \protect \BOthers {.}}{%
Parashar%
\ \protect \BOthers {.}}{%
{\protect \APACyear {2020}}%
}]{%
ParasharApJS20}
\APACinsertmetastar {%
ParasharApJS20}%
\begin{APACrefauthors}%
Parashar, T\BPBI N.%
, Goldstein, M\BPBI L.%
, Maruca, B\BPBI A.%
, us, W\BPBI H\BPBI M.%
, Ruffolo, D.%
, Bandyopadhyay, R.%
\BDBL {}Raouafi, N.%
\end{APACrefauthors}%
\unskip\
\newblock
\APACrefYearMonthDay{2020}{}{}.
\newblock
{\BBOQ}\APACrefatitle {Measures of Scale-dependent Alfv{\'{e}}nicity in the
  First {PSP} Solar Encounter} {Measures of scale-dependent alfv{\'{e}}nicity
  in the first {PSP} solar encounter}.{\BBCQ}
\newblock
\APACjournalVolNumPages{The Astrophysical Journal Supplement
  Series}{246}{2}{58}.
\newblock
\begin{APACrefDOI} \doi{10.3847/1538-4365/ab64e6} \end{APACrefDOI}
\PrintBackRefs{\CurrentBib}

\bibitem [\protect \citeauthoryear {%
Pavlova%
, Abdurashitov%
, Ulanova%
, Shushunova%
\BCBL {}\ \BBA {} Pavlov%
}{%
Pavlova%
\ \protect \BOthers {.}}{%
{\protect \APACyear {2019}}%
}]{%
Pavlova2019}
\APACinsertmetastar {%
Pavlova2019}%
\begin{APACrefauthors}%
Pavlova, O\BPBI N.%
, Abdurashitov, A\BPBI S.%
, Ulanova, M\BPBI V.%
, Shushunova, N\BPBI A.%
\BCBL {}\ \BBA {} Pavlov, A\BPBI N.%
\end{APACrefauthors}%
\unskip\
\newblock
\APACrefYearMonthDay{2019}{1}{}.
\newblock
{\BBOQ}\APACrefatitle {Effects of missing data on characterization of complex
  dynamics from time series} {Effects of missing data on characterization of
  complex dynamics from time series}.{\BBCQ}
\newblock
\APACjournalVolNumPages{Communications in Nonlinear Science and Numerical
  Simulation}{66}{}{31-40}.
\newblock
\begin{APACrefDOI} \doi{10.1016/j.cnsns.2018.06.002} \end{APACrefDOI}
\PrintBackRefs{\CurrentBib}

\bibitem [\protect \citeauthoryear {%
Peek%
\ \BBA {} Burkhart%
}{%
Peek%
\ \BBA {} Burkhart%
}{%
{\protect \APACyear {2019}}%
}]{%
peek2019}
\APACinsertmetastar {%
peek2019}%
\begin{APACrefauthors}%
Peek, J\BPBI E\BPBI G.%
\BCBT {}\ \BBA {} Burkhart, B.%
\end{APACrefauthors}%
\unskip\
\newblock
\APACrefYearMonthDay{2019}{9}{}.
\newblock
{\BBOQ}\APACrefatitle {Do Androids Dream of Magnetic Fields? Using Neural
  Networks to Interpret the Turbulent Interstellar Medium} {Do androids dream
  of magnetic fields? using neural networks to interpret the turbulent
  interstellar medium}.{\BBCQ}
\newblock
\APACjournalVolNumPages{The Astrophysical Journal}{882}{}{L12}.
\newblock
\begin{APACrefDOI} \doi{10.3847/2041-8213/ab3a9e} \end{APACrefDOI}
\PrintBackRefs{\CurrentBib}

\bibitem [\protect \citeauthoryear {%
J.~Podesta%
, Roberts%
\BCBL {}\ \BBA {} Goldstein%
}{%
J.~Podesta%
\ \protect \BOthers {.}}{%
{\protect \APACyear {2007}}%
}]{%
podesta2007}
\APACinsertmetastar {%
podesta2007}%
\begin{APACrefauthors}%
Podesta, J.%
, Roberts, D.%
\BCBL {}\ \BBA {} Goldstein, M.%
\end{APACrefauthors}%
\unskip\
\newblock
\APACrefYearMonthDay{2007}{}{}.
\newblock
{\BBOQ}\APACrefatitle {Spectral exponents of kinetic and magnetic energy
  spectra in solar wind turbulence} {Spectral exponents of kinetic and magnetic
  energy spectra in solar wind turbulence}.{\BBCQ}
\newblock
\APACjournalVolNumPages{The Astrophysical Journal}{664}{1}{543}.
\PrintBackRefs{\CurrentBib}

\bibitem [\protect \citeauthoryear {%
J\BPBI J.~Podesta%
\ \BBA {} Gary%
}{%
J\BPBI J.~Podesta%
\ \BBA {} Gary%
}{%
{\protect \APACyear {2011}}%
}]{%
Podesta_2011}
\APACinsertmetastar {%
Podesta_2011}%
\begin{APACrefauthors}%
Podesta, J\BPBI J.%
\BCBT {}\ \BBA {} Gary, S\BPBI P.%
\end{APACrefauthors}%
\unskip\
\newblock
\APACrefYearMonthDay{2011}{may}{}.
\newblock
{\BBOQ}\APACrefatitle {{MAGNETIC} {HELICITY} {SPECTRUM} {OF} {SOLAR} {WIND}
  {FLUCTUATIONS} {AS} A {FUNCTION} {OF} {THE} {ANGLE} {WITH} {RESPECT} {TO}
  {THE} {LOCAL} {MEAN} {MAGNETIC} {FIELD}} {{MAGNETIC} {HELICITY} {SPECTRUM}
  {OF} {SOLAR} {WIND} {FLUCTUATIONS} {AS} a {FUNCTION} {OF} {THE} {ANGLE}
  {WITH} {RESPECT} {TO} {THE} {LOCAL} {MEAN} {MAGNETIC} {FIELD}}.{\BBCQ}
\newblock
\APACjournalVolNumPages{The Astrophysical Journal}{734}{1}{15}.
\newblock
\begin{APACrefURL} \url{https://doi.org/10.1088/0004-637x/734/1/15}
  \end{APACrefURL}
\newblock
\begin{APACrefDOI} \doi{10.1088/0004-637x/734/1/15} \end{APACrefDOI}
\PrintBackRefs{\CurrentBib}

\bibitem [\protect \citeauthoryear {%
Qin%
, Denton%
, Tsyganenko%
\BCBL {}\ \BBA {} Wolf%
}{%
Qin%
\ \protect \BOthers {.}}{%
{\protect \APACyear {2007}}%
}]{%
Qin2007}
\APACinsertmetastar {%
Qin2007}%
\begin{APACrefauthors}%
Qin, Z.%
, Denton, R\BPBI E.%
, Tsyganenko, N\BPBI A.%
\BCBL {}\ \BBA {} Wolf, S.%
\end{APACrefauthors}%
\unskip\
\newblock
\APACrefYearMonthDay{2007}{11}{}.
\newblock
{\BBOQ}\APACrefatitle {Solar wind parameters for magnetospheric magnetic field
  modeling} {Solar wind parameters for magnetospheric magnetic field
  modeling}.{\BBCQ}
\newblock
\APACjournalVolNumPages{Space Weather}{5}{}{n/a-n/a}.
\newblock
\begin{APACrefURL} \url{http://doi.wiley.com/10.1029/2006SW000296}
  \end{APACrefURL}
\newblock
\begin{APACrefDOI} \doi{10.1029/2006SW000296} \end{APACrefDOI}
\PrintBackRefs{\CurrentBib}

\bibitem [\protect \citeauthoryear {%
Randolph-Gips%
}{%
Randolph-Gips%
}{%
{\protect \APACyear {2008}}%
}]{%
gips_2008}
\APACinsertmetastar {%
gips_2008}%
\begin{APACrefauthors}%
Randolph-Gips, M.%
\end{APACrefauthors}%
\unskip\
\newblock
\APACrefYearMonthDay{2008}{}{}.
\newblock
{\BBOQ}\APACrefatitle {A New Neural Network to Process Missing Data without
  Imputation} {A new neural network to process missing data without
  imputation}.{\BBCQ}
\newblock
\BIn{} \APACrefbtitle {2008 Seventh International Conference on Machine
  Learning and Applications} {2008 seventh international conference on machine
  learning and applications}\ (\BPG~756-762).
\newblock
\begin{APACrefDOI} \doi{10.1109/ICMLA.2008.89} \end{APACrefDOI}
\PrintBackRefs{\CurrentBib}

\bibitem [\protect \citeauthoryear {%
Rehfeld%
, Marwan%
, Heitzig%
\BCBL {}\ \BBA {} Kurths%
}{%
Rehfeld%
\ \protect \BOthers {.}}{%
{\protect \APACyear {2011}}%
}]{%
Rehfeld2011}
\APACinsertmetastar {%
Rehfeld2011}%
\begin{APACrefauthors}%
Rehfeld, K.%
, Marwan, N.%
, Heitzig, J.%
\BCBL {}\ \BBA {} Kurths, J.%
\end{APACrefauthors}%
\unskip\
\newblock
\APACrefYearMonthDay{2011}{}{}.
\newblock
{\BBOQ}\APACrefatitle {Comparison of correlation analysis techniques for
  irregularly sampled time series} {Comparison of correlation analysis
  techniques for irregularly sampled time series}.{\BBCQ}
\newblock
\APACjournalVolNumPages{Nonlinear Processes in Geophysics}{18}{}{389-404}.
\newblock
\begin{APACrefDOI} \doi{10.5194/NPG-18-389-2011} \end{APACrefDOI}
\PrintBackRefs{\CurrentBib}

\bibitem [\protect \citeauthoryear {%
Schoellhamer%
}{%
Schoellhamer%
}{%
{\protect \APACyear {2001}}%
}]{%
Schoellhamer2001}
\APACinsertmetastar {%
Schoellhamer2001}%
\begin{APACrefauthors}%
Schoellhamer, D\BPBI H.%
\end{APACrefauthors}%
\unskip\
\newblock
\APACrefYearMonthDay{2001}{8}{}.
\newblock
{\BBOQ}\APACrefatitle {Singular spectrum analysis for time series with missing
  data} {Singular spectrum analysis for time series with missing data}.{\BBCQ}
\newblock
\APACjournalVolNumPages{Geophysical Research Letters}{28}{}{3187-3190}.
\newblock
\begin{APACrefDOI} \doi{10.1029/2000GL012698} \end{APACrefDOI}
\PrintBackRefs{\CurrentBib}

\bibitem [\protect \citeauthoryear {%
Sharpe%
\ \BBA {} Solly%
}{%
Sharpe%
\ \BBA {} Solly%
}{%
{\protect \APACyear {1995}}%
}]{%
Sharpe1995}
\APACinsertmetastar {%
Sharpe1995}%
\begin{APACrefauthors}%
Sharpe, P\BPBI K.%
\BCBT {}\ \BBA {} Solly, R\BPBI J.%
\end{APACrefauthors}%
\unskip\
\newblock
\APACrefYearMonthDay{1995}{6}{}.
\newblock
{\BBOQ}\APACrefatitle {Dealing with missing values in neural network-based
  diagnostic systems} {Dealing with missing values in neural network-based
  diagnostic systems}.{\BBCQ}
\newblock
\APACjournalVolNumPages{Neural Computing and Applications}{3}{}{73-77}.
\newblock
\begin{APACrefURL} \url{http://link.springer.com/10.1007/BF01421959}
  \end{APACrefURL}
\newblock
\begin{APACrefDOI} \doi{10.1007/BF01421959} \end{APACrefDOI}
\PrintBackRefs{\CurrentBib}

\bibitem [\protect \citeauthoryear {%
Smieja%
, Łukasz Struski%
, Tabor%
, Zielínski%
\BCBL {}\ \BBA {} Spurek%
}{%
Smieja%
\ \protect \BOthers {.}}{%
{\protect \APACyear {2018}}%
}]{%
Smieja2018}
\APACinsertmetastar {%
Smieja2018}%
\begin{APACrefauthors}%
Smieja, M.%
, Łukasz Struski%
, Tabor, J.%
, Zielínski, B.%
\BCBL {}\ \BBA {} Spurek, P.%
\end{APACrefauthors}%
\unskip\
\newblock
\APACrefYearMonthDay{2018}{}{}.
\newblock
{\BBOQ}\APACrefatitle {Processing of missing data by neural networks}
  {Processing of missing data by neural networks}.{\BBCQ}.
\PrintBackRefs{\CurrentBib}

\bibitem [\protect \citeauthoryear {%
Sokolov%
\ \protect \BOthers {.}}{%
Sokolov%
\ \protect \BOthers {.}}{%
{\protect \APACyear {2013}}%
}]{%
sokolov2013}
\APACinsertmetastar {%
sokolov2013}%
\begin{APACrefauthors}%
Sokolov, I\BPBI V.%
, Van~der Holst, B.%
, Oran, R.%
, Downs, C.%
, Roussev, I\BPBI I.%
, Jin, M.%
\BDBL {}Gombosi, T\BPBI I.%
\end{APACrefauthors}%
\unskip\
\newblock
\APACrefYearMonthDay{2013}{}{}.
\newblock
{\BBOQ}\APACrefatitle {Magnetohydrodynamic waves and coronal heating: Unifying
  empirical and MHD turbulence models} {Magnetohydrodynamic waves and coronal
  heating: Unifying empirical and mhd turbulence models}.{\BBCQ}
\newblock
\APACjournalVolNumPages{The Astrophysical Journal}{764}{1}{23}.
\PrintBackRefs{\CurrentBib}

\bibitem [\protect \citeauthoryear {%
B.~Tsurutani%
\ \protect \BOthers {.}}{%
B.~Tsurutani%
\ \protect \BOthers {.}}{%
{\protect \APACyear {2002}}%
}]{%
TsurutaniGRL2002}
\APACinsertmetastar {%
TsurutaniGRL2002}%
\begin{APACrefauthors}%
Tsurutani, B.%
, Galvan, C.%
, Arballo, J.%
, Winterhalter, D.%
, Sakurai, R.%
, Smith, E.%
\BDBL {}Balogh, A.%
\end{APACrefauthors}%
\unskip\
\newblock
\APACrefYearMonthDay{2002}{}{}.
\newblock
{\BBOQ}\APACrefatitle {Relationship between discontinuities, magnetic holes,
  magnetic decreases, and nonlinear Alfv{\'e}n waves: Ulysses observations over
  the solar poles} {Relationship between discontinuities, magnetic holes,
  magnetic decreases, and nonlinear alfv{\'e}n waves: Ulysses observations over
  the solar poles}.{\BBCQ}
\newblock
\APACjournalVolNumPages{Geophysical Research Letters}{29}{11}{23--1}.
\PrintBackRefs{\CurrentBib}

\bibitem [\protect \citeauthoryear {%
B\BPBI T.~Tsurutani%
\ \protect \BOthers {.}}{%
B\BPBI T.~Tsurutani%
\ \protect \BOthers {.}}{%
{\protect \APACyear {1996}}%
}]{%
TsurutaniJGR1996}
\APACinsertmetastar {%
TsurutaniJGR1996}%
\begin{APACrefauthors}%
Tsurutani, B\BPBI T.%
, Ho, C.%
, Arballo, J.%
, Smith, E.%
, Goldstein, B.%
, Neugebauer, M.%
\BDBL {}Feldman, W.%
\end{APACrefauthors}%
\unskip\
\newblock
\APACrefYearMonthDay{1996}{}{}.
\newblock
{\BBOQ}\APACrefatitle {Interplanetary discontinuities and Alfv{\'e}n waves at
  high heliographic latitudes: Ulysses} {Interplanetary discontinuities and
  alfv{\'e}n waves at high heliographic latitudes: Ulysses}.{\BBCQ}
\newblock
\APACjournalVolNumPages{Journal of Geophysical Research: Space
  Physics}{101}{A5}{11027--11038}.
\PrintBackRefs{\CurrentBib}

\bibitem [\protect \citeauthoryear {%
B\BPBI T.~Tsurutani%
\ \BBA {} Ho%
}{%
B\BPBI T.~Tsurutani%
\ \BBA {} Ho%
}{%
{\protect \APACyear {1999}}%
}]{%
TsurutaniRGP1999}
\APACinsertmetastar {%
TsurutaniRGP1999}%
\begin{APACrefauthors}%
Tsurutani, B\BPBI T.%
\BCBT {}\ \BBA {} Ho, C\BPBI M.%
\end{APACrefauthors}%
\unskip\
\newblock
\APACrefYearMonthDay{1999}{}{}.
\newblock
{\BBOQ}\APACrefatitle {A review of discontinuities and Alfv{\'e}n waves in
  interplanetary space: Ulysses results} {A review of discontinuities and
  alfv{\'e}n waves in interplanetary space: Ulysses results}.{\BBCQ}
\newblock
\APACjournalVolNumPages{Reviews of Geophysics}{37}{4}{517--541}.
\PrintBackRefs{\CurrentBib}

\bibitem [\protect \citeauthoryear {%
B\BPBI T.~Tsurutani%
\ \protect \BOthers {.}}{%
B\BPBI T.~Tsurutani%
\ \protect \BOthers {.}}{%
{\protect \APACyear {2018}}%
}]{%
TsurutaniJGR2018}
\APACinsertmetastar {%
TsurutaniJGR2018}%
\begin{APACrefauthors}%
Tsurutani, B\BPBI T.%
, Lakhina, G\BPBI S.%
, Sen, A.%
, Hellinger, P.%
, Glassmeier, K\BHBI H.%
\BCBL {}\ \BBA {} Mannucci, A\BPBI J.%
\end{APACrefauthors}%
\unskip\
\newblock
\APACrefYearMonthDay{2018}{}{}.
\newblock
{\BBOQ}\APACrefatitle {A Review of Alfvénic Turbulence in High-Speed Solar
  Wind Streams: Hints From Cometary Plasma Turbulence} {A review of alfvénic
  turbulence in high-speed solar wind streams: Hints from cometary plasma
  turbulence}.{\BBCQ}
\newblock
\APACjournalVolNumPages{Journal of Geophysical Research: Space
  Physics}{123}{4}{2458-2492}.
\newblock
\begin{APACrefURL}
  \url{https://agupubs.onlinelibrary.wiley.com/doi/abs/10.1002/2017JA024203}
  \end{APACrefURL}
\newblock
\begin{APACrefDOI} \doi{https://doi.org/10.1002/2017JA024203} \end{APACrefDOI}
\PrintBackRefs{\CurrentBib}

\bibitem [\protect \citeauthoryear {%
Tu%
\ \BBA {} Marsch%
}{%
Tu%
\ \BBA {} Marsch%
}{%
{\protect \APACyear {1995}}%
}]{%
TuSSR95}
\APACinsertmetastar {%
TuSSR95}%
\begin{APACrefauthors}%
Tu, C.%
\BCBT {}\ \BBA {} Marsch, E.%
\end{APACrefauthors}%
\unskip\
\newblock
\APACrefYearMonthDay{1995}{}{}.
\newblock
{\BBOQ}\APACrefatitle {{MHD structures, waves and turbulence in the solar wind:
  Observations and theories}} {{MHD structures, waves and turbulence in the
  solar wind: Observations and theories}}.{\BBCQ}
\newblock
\APACjournalVolNumPages{Space Science Reviews}{73}{1}{1--210}.
\PrintBackRefs{\CurrentBib}

\bibitem [\protect \citeauthoryear {%
Usmanov%
, Matthaeus%
, Breech%
\BCBL {}\ \BBA {} Goldstein%
}{%
Usmanov%
\ \protect \BOthers {.}}{%
{\protect \APACyear {2011}}%
}]{%
UsmanovApJ11}
\APACinsertmetastar {%
UsmanovApJ11}%
\begin{APACrefauthors}%
Usmanov, A\BPBI V.%
, Matthaeus, W\BPBI H.%
, Breech, B\BPBI A.%
\BCBL {}\ \BBA {} Goldstein, M\BPBI L.%
\end{APACrefauthors}%
\unskip\
\newblock
\APACrefYearMonthDay{2011}{}{}.
\newblock
{\BBOQ}\APACrefatitle {Solar wind modeling with turbulence transport and
  heating} {Solar wind modeling with turbulence transport and heating}.{\BBCQ}
\newblock
\APACjournalVolNumPages{The Astrophysical Journal}{727}{2}{84}.
\PrintBackRefs{\CurrentBib}

\bibitem [\protect \citeauthoryear {%
van~der Holst%
\ \protect \BOthers {.}}{%
van~der Holst%
\ \protect \BOthers {.}}{%
{\protect \APACyear {2022}}%
}]{%
vanderHolstApJ22}
\APACinsertmetastar {%
vanderHolstApJ22}%
\begin{APACrefauthors}%
van~der Holst, B.%
, Huang, J.%
, Sachdeva, N.%
, Kasper, J.%
, Manchester~IV, W.%
, Borovikov, D.%
\BDBL {}others%
\end{APACrefauthors}%
\unskip\
\newblock
\APACrefYearMonthDay{2022}{}{}.
\newblock
{\BBOQ}\APACrefatitle {Improving the Alfv{\'e}n Wave Solar Atmosphere Model
  Based on Parker Solar Probe Data} {Improving the alfv{\'e}n wave solar
  atmosphere model based on parker solar probe data}.{\BBCQ}
\newblock
\APACjournalVolNumPages{The Astrophysical Journal}{925}{2}{146}.
\PrintBackRefs{\CurrentBib}

\bibitem [\protect \citeauthoryear {%
VanderPlas%
}{%
VanderPlas%
}{%
{\protect \APACyear {2018}}%
}]{%
VanderPlas2018}
\APACinsertmetastar {%
VanderPlas2018}%
\begin{APACrefauthors}%
VanderPlas, J\BPBI T.%
\end{APACrefauthors}%
\unskip\
\newblock
\APACrefYearMonthDay{2018}{5}{}.
\newblock
{\BBOQ}\APACrefatitle {{Understanding the Lomb–Scargle Periodogram}}
  {{Understanding the Lomb–Scargle Periodogram}}.{\BBCQ}
\newblock
\APACjournalVolNumPages{The Astrophysical Journal Supplement
  Series}{236}{}{16}.
\newblock
\begin{APACrefDOI} \doi{10.3847/1538-4365/aab766} \end{APACrefDOI}
\PrintBackRefs{\CurrentBib}

\bibitem [\protect \citeauthoryear {%
Velicer%
\ \BBA {} Colby%
}{%
Velicer%
\ \BBA {} Colby%
}{%
{\protect \APACyear {2005}}%
}]{%
Velicer2005}
\APACinsertmetastar {%
Velicer2005}%
\begin{APACrefauthors}%
Velicer, W\BPBI F.%
\BCBT {}\ \BBA {} Colby, S\BPBI M.%
\end{APACrefauthors}%
\unskip\
\newblock
\APACrefYearMonthDay{2005}{8}{}.
\newblock
{\BBOQ}\APACrefatitle {{A Comparison of Missing-Data Procedures for Arima
  Time-Series Analysis}} {{A Comparison of Missing-Data Procedures for Arima
  Time-Series Analysis}}.{\BBCQ}
\newblock
\APACjournalVolNumPages{Educational and Psychological
  Measurement}{65}{}{596-615}.
\newblock
\begin{APACrefDOI} \doi{10.1177/0013164404272502} \end{APACrefDOI}
\PrintBackRefs{\CurrentBib}

\bibitem [\protect \citeauthoryear {%
Verscharen%
, Klein%
\BCBL {}\ \BBA {} Maruca%
}{%
Verscharen%
\ \protect \BOthers {.}}{%
{\protect \APACyear {2019}}%
}]{%
VerscharenLRSP19}
\APACinsertmetastar {%
VerscharenLRSP19}%
\begin{APACrefauthors}%
Verscharen, D.%
, Klein, K\BPBI G.%
\BCBL {}\ \BBA {} Maruca, B\BPBI A.%
\end{APACrefauthors}%
\unskip\
\newblock
\APACrefYearMonthDay{2019}{}{}.
\newblock
{\BBOQ}\APACrefatitle {The multi-scale nature of the solar wind} {The
  multi-scale nature of the solar wind}.{\BBCQ}
\newblock
\APACjournalVolNumPages{Living Reviews in Solar Physics}{16}{1}{5}.
\PrintBackRefs{\CurrentBib}

\bibitem [\protect \citeauthoryear {%
Wu%
\ \protect \BOthers {.}}{%
Wu%
\ \protect \BOthers {.}}{%
{\protect \APACyear {2013}}%
}]{%
wu_2013}
\APACinsertmetastar {%
wu_2013}%
\begin{APACrefauthors}%
Wu, P.%
, Perri, S.%
, Wan, M.%
, Matthaeus, W\BPBI H.%
, Shay, M\BPBI A.%
, Goldstein, M\BPBI L.%
\BDBL {}Chapman, S.%
\end{APACrefauthors}%
\unskip\
\newblock
\APACrefYearMonthDay{2013}{}{}.
\newblock
{\BBOQ}\APACrefatitle {Intermittent heating in solar wind and kinetic
  simulations} {Intermittent heating in solar wind and kinetic
  simulations}.{\BBCQ}
\newblock
\APACjournalVolNumPages{The Astrophysical Journal Letters}{763}{}{}.
\PrintBackRefs{\CurrentBib}

\bibitem [\protect \citeauthoryear {%
Yi%
, Moon%
, Lim%
, Park%
\BCBL {}\ \BBA {} Lee%
}{%
Yi%
\ \protect \BOthers {.}}{%
{\protect \APACyear {2021}}%
}]{%
Yi2021}
\APACinsertmetastar {%
Yi2021}%
\begin{APACrefauthors}%
Yi, K.%
, Moon, Y\BHBI J.%
, Lim, D.%
, Park, E.%
\BCBL {}\ \BBA {} Lee, H.%
\end{APACrefauthors}%
\unskip\
\newblock
\APACrefYearMonthDay{2021}{mar}{}.
\newblock
{\BBOQ}\APACrefatitle {Visual Explanation of a Deep Learning Solar Flare
  Forecast Model and Its Relationship to Physical Parameters} {Visual
  explanation of a deep learning solar flare forecast model and its
  relationship to physical parameters}.{\BBCQ}
\newblock
\APACjournalVolNumPages{The Astrophysical Journal}{910}{1}{8}.
\newblock
\begin{APACrefURL} \url{https://doi.org/10.3847/1538-4357/abdebe}
  \end{APACrefURL}
\newblock
\begin{APACrefDOI} \doi{10.3847/1538-4357/abdebe} \end{APACrefDOI}
\PrintBackRefs{\CurrentBib}

\bibitem [\protect \citeauthoryear {%
Zhao%
, Lange%
\BCBL {}\ \BBA {} Meissner%
}{%
Zhao%
\ \protect \BOthers {.}}{%
{\protect \APACyear {2020}}%
}]{%
Zhao2020}
\APACinsertmetastar {%
Zhao2020}%
\begin{APACrefauthors}%
Zhao, J.%
, Lange, H.%
\BCBL {}\ \BBA {} Meissner, H.%
\end{APACrefauthors}%
\unskip\
\newblock
\APACrefYearMonthDay{2020}{5}{}.
\newblock
{\BBOQ}\APACrefatitle {Gap-filling continuously-measured soil respiration data:
  A highlight of time-series-based methods} {Gap-filling continuously-measured
  soil respiration data: A highlight of time-series-based methods}.{\BBCQ}
\newblock
\APACjournalVolNumPages{Agricultural and Forest Meteorology}{285-286}{}{}.
\newblock
\begin{APACrefDOI} \doi{10.1016/J.AGRFORMET.2020.107912} \end{APACrefDOI}
\PrintBackRefs{\CurrentBib}

\end{thebibliography}

%
%
%
%
%

\end{document}